\documentclass[aps,prd,twocolumn,amsmath,amssymb,amsfonts,nofootinbib,superscriptaddress,
altaffilletter,10pt]{revtex4-1}

\usepackage{graphicx}
\usepackage{dcolumn}
\usepackage{bm}
\usepackage{url}
\usepackage{amssymb}
\usepackage{amsmath}
\usepackage{longtable}
\usepackage{rotating}
\usepackage{color}
\usepackage{epsfig}
\usepackage{epsf}
\usepackage{fancyhdr}
\usepackage{supertabular}
\usepackage{hyperref}
\usepackage[raggedright,hang]{subfigure}
\usepackage[usenames,dvipsnames]{xcolor}
\usepackage{marginnote}

\newcommand{\GCSearch}{GC search}
\newcommand{\ba}{\begin{align}}
\newcommand{\ea}{\end{align}}

\newcommand{\be}{\begin{equation}}
\newcommand{\ee}{\end{equation}}
\newcommand{\F}{\mathcal{F}}

\newcommand{\seg}{\text{seg}}
\newcommand{\Hz}{\text{Hz}}
\newcommand{\s}{\text{s}}
\newcommand{\ys}{\text{yr}}
\newcommand{\yr}{\text{yr}}

\newcommand{\Max}{\text{max}}
\newcommand{\Min}{\text{min}}
\newcommand{\satf}{\langle2\mathcal{F}\rangle}
\newcommand{\R}{\mathcal{R}}
\newcommand{\cC}{\mathcal{C}}
\newcommand{\cB}{\mathcal{B}}
\newcommand{\cA}{\mathcal{A}}
\newcommand{\cD}{\mathcal{D}}
\newcommand{\cS}{\mathcal{S}}
\newcommand{\sft}{\text{sft}}
\newcommand{\cand}{_{\text{cand}}}
\newcommand{\coh}{_{\text{coh}}}
\newcommand{\SFT}{\text{SFT}}
\newcommand{\Sh}{S_{\mathrm{h}}}

\begin{document}

\title{Postprocessing methods used in the search for continuous gravitational-wave signals from the
Galactic Center}

\author{Berit Behnke}
\email{berit.behnke@aei.mpg.de}
\affiliation{Max-Planck-Institut f\"ur Gravitationsphysik, \\
Am M\"uhlenberg 1, 14476 Postdam and Callinstrasse 38, 30167 Hannover, Germany}
\author{Maria Alessandra Papa}
\email{maria.alessandra.papa@aei.mpg.de}
\affiliation{Max-Planck-Institut f\"ur Gravitationsphysik, \\
Am M\"uhlenberg 1, 14476 Postdam and Callinstrasse 38, 30167 Hannover, Germany}
\affiliation {University of Wisconsin-Milwaukee, Milwaukee, Wisconsin 53201, USA }
\author{Reinhard Prix}
\affiliation{Max-Planck-Institut f\"ur Gravitationsphysik \\
Callinstrasse 38, 30167 Hannover, Germany}
\date{\today}

\begin{abstract}
The search for continuous gravitational-wave signals
requires the development of
techniques that can effectively explore the low-significance regions of the candidate set.
In this paper we present the methods that were developed for a search for continuous
gravitational-wave signals from the Galactic Center \cite{myobspaper}. First, we present a
data-selection method that
increases the
sensitivity of the chosen data set by 20\%-30\% compared to the selection methods used in
previous
directed searches. Second, we introduce postprocessing methods that
reliably rule out candidates that stem from random fluctuations or disturbances in the
data. In the context of [J. Aasi {\it et al.}, Phys.~Rev.~D 88, 102002 (2013)] their use enabled the
investigation of
marginal candidates (three
standard deviations below the loudest expected candidate in Gaussian noise from the entire
search). Such low-significance regions had not been explored in continuous gravitational-wave
searches
before. We finally present a new procedure for deriving upper limits on the gravitational-wave
amplitude, which
is several times faster with respect to the standard injection-and-search approach commonly used.
\end{abstract}


\maketitle
\newpage


\section{Introduction}


The ultimate goal of every gravitational-wave (GW) search is the confident detection of a signal in
the data. For most
searches, when the initial analysis has not produced a significant candidate that can be
confirmed as a confident detection, no further postprocessing can change this result. This is mainly
the case
for searches for transient signals. In contrast, for a {\it continuous} gravitational-wave (CW)
signal the scientific potential of the data is
not yet exhausted. Since a CW signal is present during the whole
observation time, a more sensitive follow-up search on interesting
candidates can be performed.

One of the factors that influences the final sensitivity of the whole search is
the finite number of candidates that one can follow up
with limited computational resources. An effective way to reduce the number of unnecessary
follow-ups, and hence increase the
sensitivity of the search, is to develop techniques that identify spurious low-significance
candidates generated by common artifacts in the data.


The first CW search that systematically explored this low-significance region was the search
\cite{myobspaper} (from now on we will refer to it as the ``\GCSearch{}'') for CW signals from
isolated rotating compact objects at the Galactic Center (GC). We refer the reader to
\cite{myobspaper} and references therein for the astrophysical
motivation of such a search. While that paper focuses on the observational results, here
we present the studies that support and characterize the postprocessing techniques
developed to yield those search results. Such techniques have allowed the inspection of marginal
candidates (i.e. with
significances about three standard deviations below the expected significance of the maximum of the
 entire search in Gaussian noise). Even for such low-significance candidates we were able to
discern whether they were disturbances or random fluctuations, or whether they were worth
further investigation.  

We also present in this paper two further techniques, first used in the \GCSearch{}. One is a
new and
computationally efficient method to determine frequentist loudest-event upper limits. Our method is
several times faster
than the standard method used for many CW searches. The other is a
data-selection
criterion. We
compare our method with three
different data-selection approaches from the literature and illustrate the gain in
sensitivity of our selection method.

This paper is structured as follows: we start in Sec.~\ref{lab:thesearch} with a short summary of
the setup of the \GCSearch{} and illustrate the data-selection criterion. In
Sec.~\ref{lab:postprocessing} we describe the different postprocessing steps which include a
relaxed method (with respect to previously used methods) to clean the data from known artifacts
(Sec.~\ref{lab:knownlines}); the clustering
of candidates that
can be ascribed to the same origin (Sec.~\ref{lab:clustering}); the $\F$-statistic consistency veto
that
checks for consistent analysis results from different detectors (Sec.~\ref{lab:consistency}); the
selection of a significant subset of candidates (Sec.~\ref{lab:significanceThreshold}); a
veto based on the expected permanence of continuous GW signals (Sec.~\ref{lab:permanence}); and a
coherent follow-up search that identifies CW candidates with high confidence
(Sec.~\ref{lab:followup}). In Sec.~\ref{lab:upperlimits} we present a semi analytic procedure
to compute the frequentist loudest-event GW amplitude upper limits at highly reduced computational
cost. In Sec.~\ref{lab:secondspindown} we discuss the detection efficiency
for an additional astrophysically relevant class of signals, which are not the target population for
which the search was originally developed: namely, signals with a second-order frequency
derivative.
The paper ends with a discussion of the results in Sec.~\ref{lab:discussion}.

%
%
\section{The search}
\label{lab:thesearch}

\subsection{The parameter space}
The \GCSearch{} aimed to detect CW signals from the Galactic Center by
searching for different CW wave shapes (i.e.\ different signal templates). These are defined
by different frequency and spin-down (time derivative of the frequency) values and by a single
sky position corresponding to the GC.

The search employed a semicoherent stack-slide method\footnote{Implemented as the code
  \texttt{lalapps\_HierarchSearchGCT} in the LIGO Algorithm Library (LALSuite) \cite{lalsuite}.}:
630 data segments of
length $T_\seg=11.5$\,h are separately analyzed and the results are afterwards combined. The
coherent
analysis of each data segment is performed for every template using a matched filtering method
\cite{jks,Cutler:2005hc,Prix:2006wm}. The resulting detection statistic $2\F$ is a measure of how
much more likely it is that a CW signal with the template parameters is present in the
data rather than just Gaussian noise. If the detector noise is Gaussian, the $2\F$ values follow a
$\chi^2$ distribution with four
degrees of freedom and a noncentrality parameter that equals the squared signal-to-noise ratio
(SNR),
$\rho^2$ [see Eq.~70 of \cite{jks}]. The $\F$-statistic values are combined using
techniques described in \cite{Brady1998b,PhysRevD.70.082001,Pletsch2009}. The result is an average
value $\satf$ for each template, where the angle brackets denote the average over segments. The
combination of the template parameters $\{f, \dot f, \alpha,
\delta\}$ and $\satf$ will be referred to as a \textit{candidate}. \par

The sky coordinates we targeted are those of Sagittarius~A* (Sgr~A*) which we use as synonym for the
GC. Since there are no specific sources with known rotation frequency and spin-down to target at the
GC,
the \GCSearch{} covered a large range in frequency and spin-down:
\be 78\,\Hz\leq f\leq 496\,\Hz \ee
in frequency $f$ and
\be 0  \leq -\dot f\leq \frac{f}{200\,\yr}\ee
in spin-down $\dot f$.

These ranges are tiled with discrete template banks which are different between the coherent and
incoherent stages.
The incoherent combination is performed on a grid that is finer in spin-down than the grid used
for the coherent searches. The frequency grid is the same for both stages. The spacings of the
template grids were:
\begin{align} \label{eq:grid-spacings}
&\delta f = \frac{1}{T_\text{seg}} = 2.4\times10^{-5}\,\Hz,\nonumber \\
&\delta \dot f_\text{coarse} = \frac{1}{T_\text{seg}^2} = 5.8\times10^{-10}\,\Hz/\s,\ \
\text{and}\nonumber \\
&\delta \dot f_\text{fine} = \frac{1}{\gamma T_\text{seg}^2} = 1.8\times10^{-13}\,\Hz/\s,
\end{align}
where $\gamma=3225$ is the refinement factor (following \cite{Pletsch2009}. \par

A GW signal will in general have parameters that lie between the template grid
points. This gives rise to a loss in detection efficiency with respect to a search done with an
infinitely fine template grid. The mismatch $m$ is defined as the fractional loss in SNR$^2$
due to the offset between the actual signal and the template parameters:
\be\label{eq:mismatch}
m = \frac{\rho^2_\text{perfect match} - \rho^2_\text{mismatched}}{\rho^2_\text{perfect match}},
\ee
where $\rho^2_\text{perfect match}$ is the SNR$^2$ obtained for a template that has exactly the
signal
parameters and $\rho^2_\text{mismatched}$ is the SNR$^2$ obtained from a template whose parameters
are
mismatched with respect to those of the signal.

The mismatch distribution for the grid spacings given in Eq.~\ref{eq:grid-spacings} was obtained by
a
Monte Carlo study in which $5\,000$ realizations of fake data are created,\footnote{The fake data
are
created with a code called \texttt{lalapps\_Makefakedata\_v4} which is also part of
\cite{lalsuite}.} each containing a CW signal (and no noise) with uniformly randomly distributed
signal parameters (frequency $f$, spin-down $\dot f$, intrinsic phase $\phi_0$, polarization angle
$\psi$ and cosine of the inclination angle,
$\cos\iota$) within the searched parameter space and right ascension and declination values
randomly
distributed within a disk of radius $R=10^{-3}$\,rad around Sgr~A*. The data set used in the
\GCSearch{} and the fake data have the same timestamps.

\begin{figure}
   \centering
   \includegraphics[trim = 20mm 0mm 5mm 0mm, clip, width=0.5\textwidth]{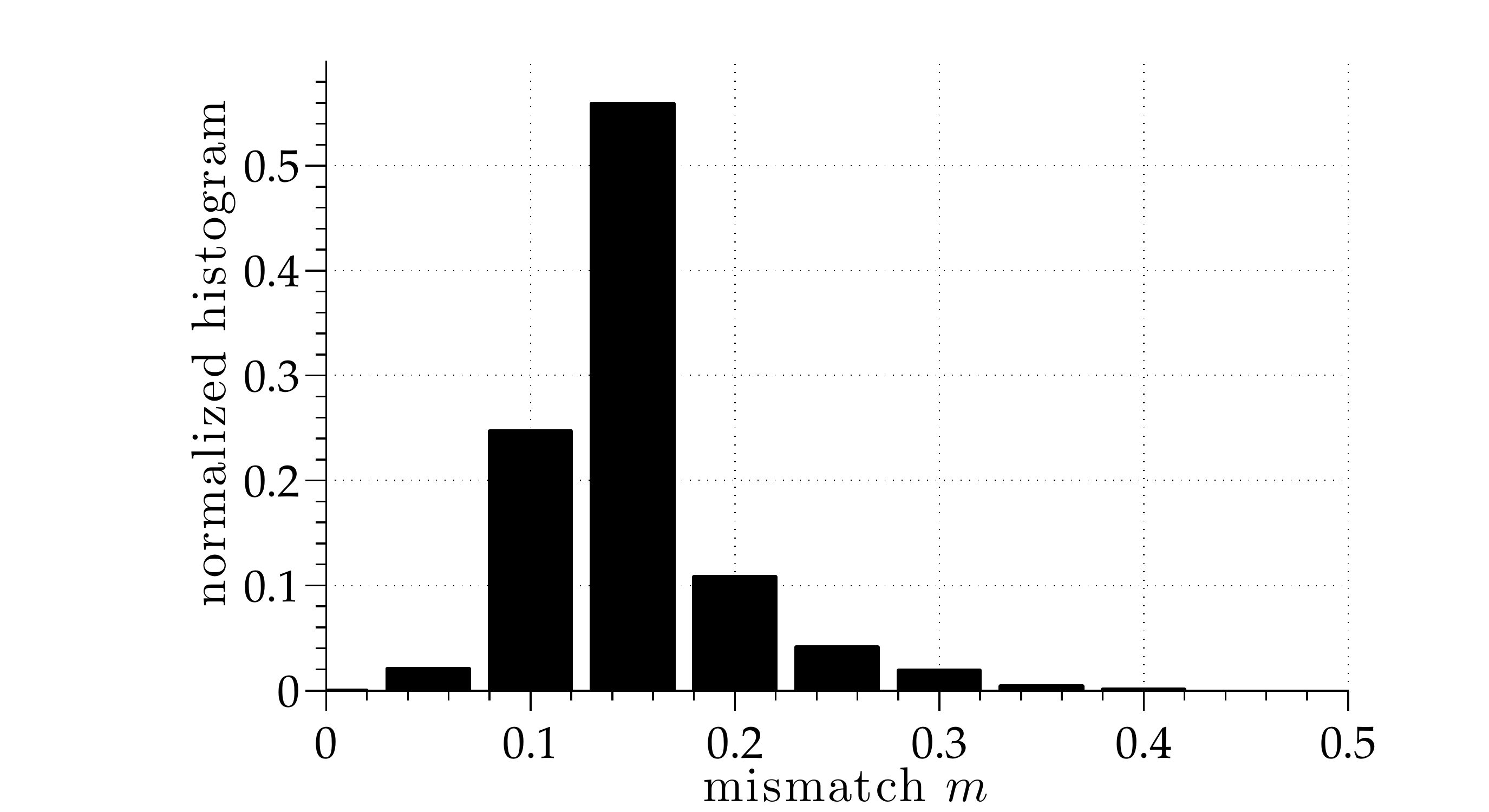}
   \caption{The histogram shows the mismatch distribution for the search template grid of
Eq.~\ref{eq:grid-spacings}. The average mismatch is $\langle m\rangle \approx 15\%$.}
   \label{fig:totalmismatch}
\end{figure}

The fake data are then analyzed targeting the sky location of Sgr~A* with the original search
template grid in $f$ and $\dot f$, restricted to a small ($100\times100$ grid points) region around
the injection, and the
largest value $\rho^2_\text{mismatched}$ is identified.
A second analysis is performed targeting the exact injection parameters to obtain
$\rho^2_\text{perfect match}$. Figure~\ref{fig:totalmismatch} shows
the normalized distribution of the $5\,000$ mismatch
values that are obtained with the described procedure. The average mismatch is found to be
$\langle m\rangle \approx 0.15$; only in a small
fraction of cases ($\lesssim 1\%$) can the loss be as high as $40\%$. \par

The parameter space searched by the \GCSearch{} comprises $\sim 4.4\times 10^{12}$
templates and is split into 10\,678 smaller compute jobs, each of which reports the most significant
100\,000
candidates, yielding a total of $10^9$ candidates to postprocess.

\subsection{Comparison to metric grid construction}
\label{sec:comp-metr-templ}

We can try to estimate the expected mismatch for this template bank using the analytic phase metric
[e.g. see Eq.(10) in \cite{Wette2008}] and combining fine- and coarse-grid
mismatches (assuming a $\mathbb{Z}_2$-like grid structure) by summing them according to Eq.(22)
of \cite{PrixShaltev:2012}.
This naive estimate, however, yields an average mismatch of
$\langle m\rangle \approx 0.56$, overestimating the measured value by nearly a factor of 4. One
reason for this is the well-known inaccuracy of the metric approximation (particularly the
phase metric) for short coherent-search times $\lesssim \mathcal{O}(\text{days})$; see
\cite{Prix:2006wm,Wette:2013wza}.
The second reason stems from the nonlinear effects that start to matter for mismatches $\gtrsim 0.2$
(e.g.\ see Fig.~10 in \cite{Prix:2006wm} and Fig.~7 in \cite{Wette:2013wza}).
A more detailed metric simulation of this template bank, using the average-$\F$ metric (instead of
phase metric) and folding in the empirical nonlinear behavior puts the expected average mismatch at
$\langle m\rangle \approx 0.2$.

Despite the quantitative discrepancy of the simple phase metric, using it to guide template-bank
construction (followed by Monte Carlo testing) would still have been useful.
For example, the simplest metric grid construction of a square lattice aiming for an
average mismatch of $\sim 0.15$ would result in grid spacings
\begin{equation} \label{eq:grid-spacings-ALT1}
\delta f' = \frac{0.37}{T_{\text{seg}}}\,,\;\;
\delta \dot f'_\text{coarse} = \frac{2.86}{T_{\text{seg}}^2}\,,
\end{equation}
and $\delta \dot f'_{\text{fine}} = \delta \dot f'_{\text{coarse}} / \gamma$. This corresponds to
a $\sim6\%$ reduction in computing cost compared to the original search, and results in a
measured mismatch distribution with an average $\langle m\rangle \approx 0.09$, which corresponds to
an improvement in sensitivity. Relaxing those spacings by a factor $1.7$, namely
\begin{equation} \label{eq:grid-spacings-ALT3}
\delta f'' = \frac{0.63}{T_{\text{seg}}}\,,\;\;
\delta \dot f''_\text{coarse} = \frac{4.87}{T_{\text{seg}}^2}\,,
\end{equation}
and $\delta \dot f''_{\text{fine}} = \delta \dot f''_{\text{coarse}} / \gamma$, results in a
measured average mismatch of $\langle m\rangle \approx 0.14$, i.e.\ similar to the original setup,
but at
a total computing cost reduced by about a factor of $\sim3.1$.
Such large gains are possible here due to the fact that the original search
grid of Eq.~\eqref{eq:grid-spacings} deviates strongly from a square lattice in the metric sense:
namely the frequency spacing is about $\sim 80$ times larger in mismatch than the spin-down spacing.

%
%
\subsection{Data selection}
\label{lab:dataselection}

The data used in the \GCSearch{} come from two of the three initial LIGO detectors,
H1 and L1, and were collected at the time of the fifth science run (S5)\footnote{The fifth science
run started on November~4, 2005 at 16:00~UTC in
Hanford and on November~14, 2005 at 16:00~UTC in Livingston and ended on October~1, 2007 at
00:00~UTC.}
\cite{Abbott2009e}. The recorded data were calibrated to produce a GW strain $h(t)$ time series
\cite{Abbott2009d, Abbott2009e, Abadie:2010px} which was then broken into $T_{\sft}=1800\,\s$ long
stretches,
high-pass filtered above 40\,Hz, Tukey windowed, and Fourier transformed to form short
Fourier transforms (SFTs) of $h(t)$.

Based on constraints stemming from the available computing power, 630 segments, each spanning
$T_\seg=11.5$~h and generally comprising data from both detectors, were used in the \GCSearch{}.
These
data
cover a period of 711.4 days which is 98.6\% of the total duration of
S5. The segments were not completely filled with data from both detectors: the total amount of
data (from both detectors) corresponds to 447.1 days, which means an average fill level per detector
per segment of $\sim74\%$.

Different approaches exist for selecting the segments to search from a given data set. We compare
different selection criteria by studying the sensitivity of different sets of S5 segments chosen
according to the different criteria. We start by creating a set $\cS$ comprising all possible
segments of fixed length. We then pick 630 nonoverlapping segments from this set according to each
criterion.

Set $\cS$ is constructed as follows: each segment covers a time
$T_\seg$ and neighboring segments overlap each other by $T_\seg$ - 30~min. The first
segment starts at the time\-stamp of the first SFT of S5. All SFTs that lie entirely within $T_\seg$
are assigned to
that segment. The second segment starts half an hour later, and so on.

The different sets of segments are constructed as follows: every criterion assigns a different
figure of merit to each segment in $\cS$. For each
criterion we then select the
segment with the highest value of the corresponding figure of merit while all overlapping
segments are removed from $\cS$. From the remaining list, the segment
with the next-highest value of the figure of merit is selected, and again all overlapping segments
are
removed. This is repeated until 630 segments are selected. Since we consider four different criteria
this procedure generates four different
sets of segments labeled $\cA$, $\cB$, $\cC$ and $\cD$. The four different criteria reflect
data-selection
choices made in past searches, including the one
specifically developed for the \GCSearch{}, which we describe in the following.

The figures of merit are computed
at a single fixed frequency. This
is possible because the relative performance of the different selection criteria in Gaussian noise
does not depend on the actual value of the frequency of the signal. Hence we choose 150\,Hz
which is in a spectral area that does not contain known disturbances and is representative of the
noise in the large majority of the data set.

The first selection criterion is solely based on the fill level of data
per segment,
as was done in \cite{Abbott2009a, Abbott2009d}.
The corresponding figure of merit
is the number of SFTs in each segment, namely
\begin{equation}
  \label{eq:3}
  Q_\cA = N_\seg^\SFT\,.
\end{equation}
The resulting data set is referred to as set $\cA$.

The second criterion was used in the first coherent search for CWs from the central
object in the supernova remnant Cassiopeia A, where the data were selected based on the
noise level of the data in addition to the fill level \cite{Wette2010}.
The figure of merit is defined as the harmonic sum in each segment of the noise
power spectral density at 150\,Hz $\Sh$ \cite{Wette2010}:
\begin{equation}
  \label{eq:1}
  Q_\cB =\sum_{k=1}^{N_\seg^\SFT} \frac{1}{\Sh^k}\,.
\end{equation}
The sum is over all SFTs in the appropriate segment. The resulting data set is referred to as set
$\cB$.

The third criterion is the first to take into account not only the amount of data
in each segment, $N_\seg^\SFT$, and the quality of the data, expressed in terms of the
strain noise $\Sh$, but also the quality
of the data, expressed in terms of the strain noise $\Sh$, and the sensitivity of the
detector network to signals from a certain sky position at the time when the data were recorded.
This
criterion was used
for the fully coherent search for CW signals from Scorpius X-1, using 6\,h of data from LIGO's
second science run [Eq.~36 in
\cite{Abbott2007a}]. We define an equivalent figure of merit as
\begin{equation}\label{figureofmerit}
  Q_\cC = N_\seg^\SFT \frac{\sum_{k=1}^{N_\seg^\SFT}\mathcal{P}^k}{\sum_{k=1}^{N_\seg^\SFT} \Sh^k},
\end{equation}
where $\mathcal{P}^k=(F_+^k)^2 + (F_\times^k)^2$ depends on the antenna pattern functions ${F_+^k}$
and
${F_\times^k}$ \cite{jks}, computed at the midtime of each SFT $k$ in the considered segment. The
data
set that we obtain with this method is referred to as set $\cC$.

The fourth criterion is the one used in the \GCSearch{}. The figure of merit is the average expected
detection statistic:
\begin{equation}
  \label{eq:2}
  Q_\cD = \overline{\mathrm{E}[2\F]}\,,
\end{equation}
where the average denoted by $\overline{\protect\phantom{{}2F{}}}$ is over a signal population with
a
fixed GW amplitude $h_0$ and random polarization parameters ($\cos\iota$ and $\psi$) and where
$\mathrm{E}[]$ denotes the expectation value over noise realizations. The resulting data set is
referred to as set $\cD$.

\begin{figure}
\centering
\includegraphics[trim = 3mm 0mm 5mm 0mm, clip,
width=0.5\textwidth]{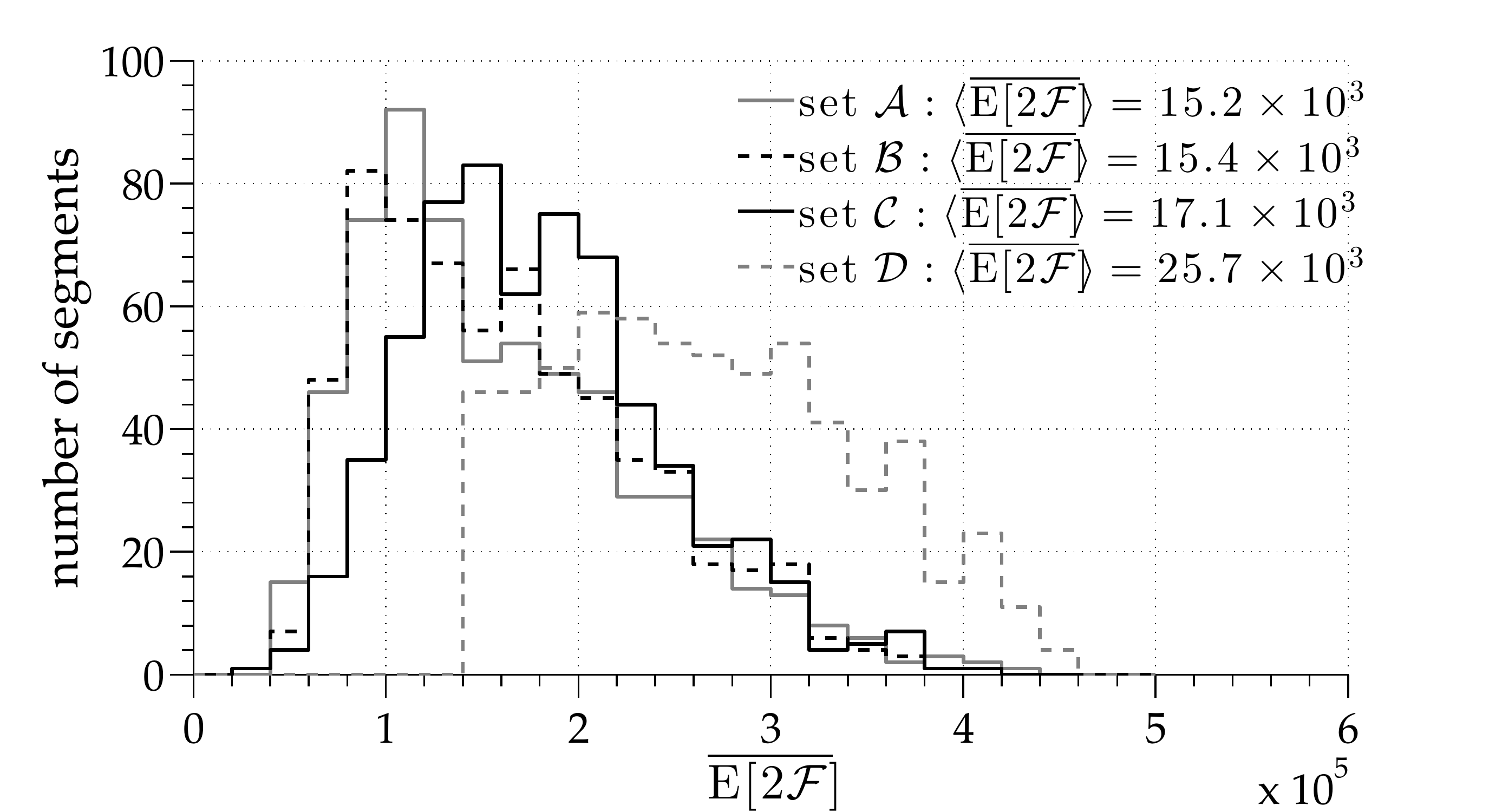}
\caption{The distributions of the average expected detection statistic over segments,
$\overline{\mathrm{E}[2\F]}$. $\overline{\protect\phantom{{}2F{}}}$ denotes the average over a
population of signals with a fixed fiducial GW amplitude and random polarization parameters
($\cos\iota$ and $\psi$).
The figure shows
that a large fraction of the segments in set $\cD$
has high $\overline{\mathrm{E}[2\F]}$ values compared to the other sets. We also compare for each
set the average value of $\overline{\mathrm{E}[2\F]}$ over the 630 segments,
$\langle\overline{\mathrm{E}[2\F]}\rangle$. The
higher that number, the more sensitive the data set is.}
\label{fig:datacomparison}
\end{figure}

In order to compare the sensitivity of searches carried out on these different data sets we compute
the expected
$\F$-statistic value for a signal coming from the GC for the single segments in each
of the data sets. Figure~\ref{fig:datacomparison} shows distributions of the average value of the
expected $2\F$ for a
population of sources at the GC with a fixed fiducial GW amplitude (and averaged over the
polarization parameters $\cos\iota$ and $\psi$) for each segment in sets
$\cA,~\cB,~\cC$ or $\cD$. The integral of each histogram is the number of segments, 630. By
construction,
sets $\cA$, $\cB$ and $\cC$ cannot be more sensitive than $\cD$, but here we quantify the
differences:
The data set used for
the \GCSearch{} ($\cD$) results in an average $2\F$ value (over segments and over a population of
signals) that is about 1.5 times higher than that of set $\cC$. This translates into a
gain of roughly 20\%-30\% in minimum detectable GW amplitude for set $\cD$ with respect to set
$\cC$.
The ratio of the average $2\F$ value between sets $\cC$, $\cB$ and $\cA$ does not exceed 1.1.

\begin{figure}
\centering
\includegraphics[trim = 3mm 0mm 5mm 0mm, clip,
width=0.5\textwidth]{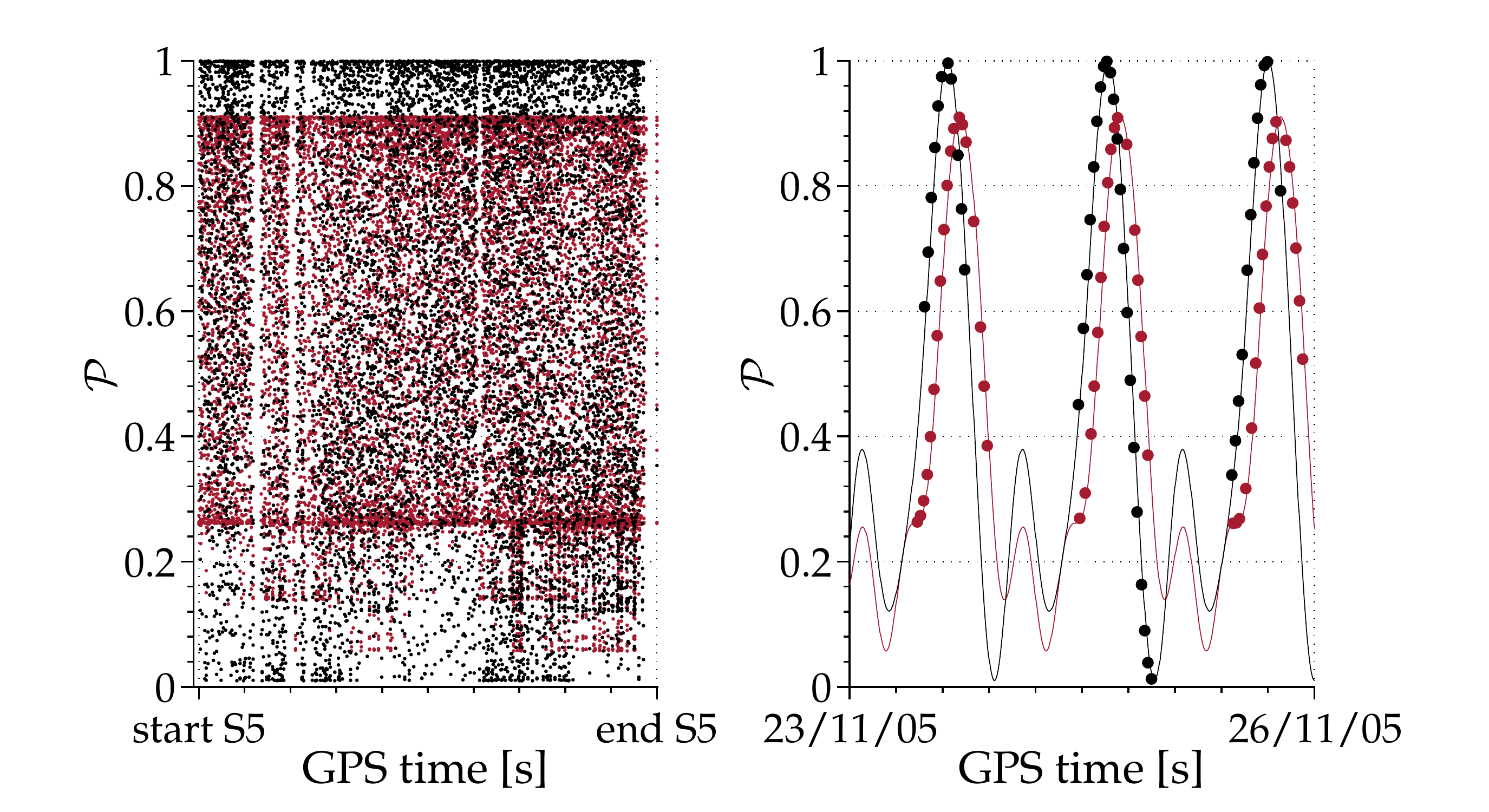}
\caption{The antenna pattern functions $\mathcal{P}(t_k)=F_+(t_k)^2 + F_\times (t_k)^2$ at the
midtime
$t_k$ of each SFT of set
$\cD$ ($k$ indicates the order number of the SFTs and $t_k$ is on the {\it x} axis). Red color shows
the values for H1, black those for L1. The right plot is a zoom of
the
left plot for a duration of $2.9$~days within the first weeks of S5. As described above, the
segments are selected to maximize the
expected $2\F$ value which explains the distribution of points along the solid curves. The maximum
and minimum values of $\mathcal{P}(t)$ for the Hanford detector, just above 0.9 and just below
0.08, are due to the fact that at the latitude of Hanford ($46^\circ 17^\prime 8^{\prime\prime}$)
the GC (which has a declination of $-29^\circ 0^\prime 28^{\prime\prime}$) can never reach the
zenith.}
\label{fig:SegmentComparisonDetectorState}
\end{figure}

An interesting property of the selected data in set $\cD$ is that the segments were recorded at
times when
the
detectors had especially favorable orientation with respect to the GC. We can clearly see this from
the zoom of Fig.~\ref{fig:SegmentComparisonDetectorState} and by comparing the average antenna
patterns $\overline{\mathcal{P}(\mathcal{X})} = (1/N^{\mathcal{X}}_{\sft})\sum_{k\in
\mathcal{X}}\mathcal{P}_k$ for the SFTs $k$ in
the data sets $\mathcal{X} = \cA,\,\cB,\,\cC,\,\cD$:
\be\label{eq:fplusfcrosscomparison}
\overline{\mathcal{P}(\cA)} = 0.38, ~~\overline{\mathcal{P}(\cB)} = 0.37,
\overline{\mathcal{P}(\cC)} = 0.50, ~~\overline{\mathcal{P}(\cD)} = 0.56.
\ee

\section{Postprocessing}
\label{lab:postprocessing}

%
%
\subsection{Known-lines cleaning}
\label{lab:knownlines}

Terrestrial disturbances affect GW searches in undesired ways. Some disturbances
originate from the detectors themselves, like the ac power line harmonics or some mechanical
resonances of the
apparatus. These (quasi)stationary spectral lines affect the analysis results by generating
suspiciously
large $\satf$
values. Over the last years, knowledge about noise sources has been collected, mostly
by searching for correlations between the GW channel of the detector and other auxiliary channels.
However, only rarely can those noise sources be mitigated and hence we need to deal with candidates
that
we suspect are due to disturbances. One of the approaches taken so far consisted of removing
candidates that lie within disturbed frequency bands.
This can happen both by excluding up front certain frequency bands from the search or by so-called
``known-lines cleaning procedures,'' two variants of which are discussed below.

\subsubsection{The ``strict'' known-lines cleaning}

This cleaning procedure, as it has been used in past all-sky searches (see, for example
\cite{Aasi:2012fw}),
removes all candidates whose value of the detection statistic may have had contributions from data
contaminated by known disturbances. In order to determine whether data from frequency bins in a
corrupted band have contributed to the detection statistic at a given parameter-space point
$\{f,\dot f\}$, one has to determine the frequency evolution of the GW signal as observed at the
detector during the observation time and see if there is any overlap with the disturbed band. If
there is, the candidate from that parameter-space point is not considered. The method actually used
is
even coarser, since the sky position of the candidate is ignored and the span of the
instantaneous frequency is assumed to be the absolute maximum possible over a year, independently of
sky position.

Had this method been applied in the \GCSearch{}, the large spin-down values searched and the
regular occurrence of the $1\,\Hz$ harmonics in the S5 data (see Tables VI and VII in
\cite{Aasi:2012fw}) would have resulted in a huge loss of $\sim 88.6\%$ of all candidates.
Such loss is unnecessary as is shown below, and a more relaxed variation on this veto scheme
can be used.

\subsubsection{The ``relaxed'' known-lines cleaning}
\label{ch:flexibleknownlinescleaning}

The reason for the above-mentioned loss is twofold: the regular occurrence of 1\,Hz harmonics and
the large spin-down values searched in the \GCSearch{}. If a signal has a large spin-down magnitude
$|\dot f|$, its
intrinsic frequency changes rapidly over time and hence it ``sweeps'' quickly through a large range
in
frequency. Most templates have spin-down values large enough to enter one of the 1\,Hz
frequency
bands at some point, but they also sweep through the contaminated bands quickly. Consider for
instance a contaminated band from the $1\,\Hz$ harmonic at $496\,\Hz$, where the intrinsic width of
the $1\,\Hz$ harmonic is maximal and reaches $\Delta f =
0.08\,\Hz$. A signal with a spin-down of magnitude $|\dot f_\text{av}| = 5.9\times 10^{-8}\,\Hz/\s$
(the average spin-down value searched in the \GCSearch{}) sweeps through such a band in
$\Delta t = \Delta f / |\dot f_\text{av}| \simeq 16$\,days.
Since one data segment spans only about half a day, this means that at most about 32 of the
630 segments contribute data to the detection
statistic that may be affected by that artifact, which is only about $\sim 5\%$. This means that
only a very
small percentage of the data used for the analysis of high spin-down templates is potentially
contaminated by a $1\,\Hz$ line.

Based on this, we relax the known-lines cleaning procedure and allow for a certain
amount of data to be potentially contaminated by a 1\,Hz harmonic. Since the 1\,Hz lines are the
only
artifacts with such a major impact on the number of surviving candidates, the other known spectral
lines are still vetoed ``strictly.'' In the \GCSearch{}, all candidates with frequency and spin-down
values such that no more than $30\%$ of the data used for the analysis of the candidate are
potentially contaminated by a 1\,Hz line are kept. One could easily argue for a larger threshold
than
$30\%$, given the negligible impact of the spectral lines. As a measure of safety, all candidates
which pass this procedure only due to the relaxation of the line cleaning are labeled. Further
investigations can then fold in that information, if necessary.

\begin{figure}[]
    \centering
	\includegraphics[width=0.5\textwidth]{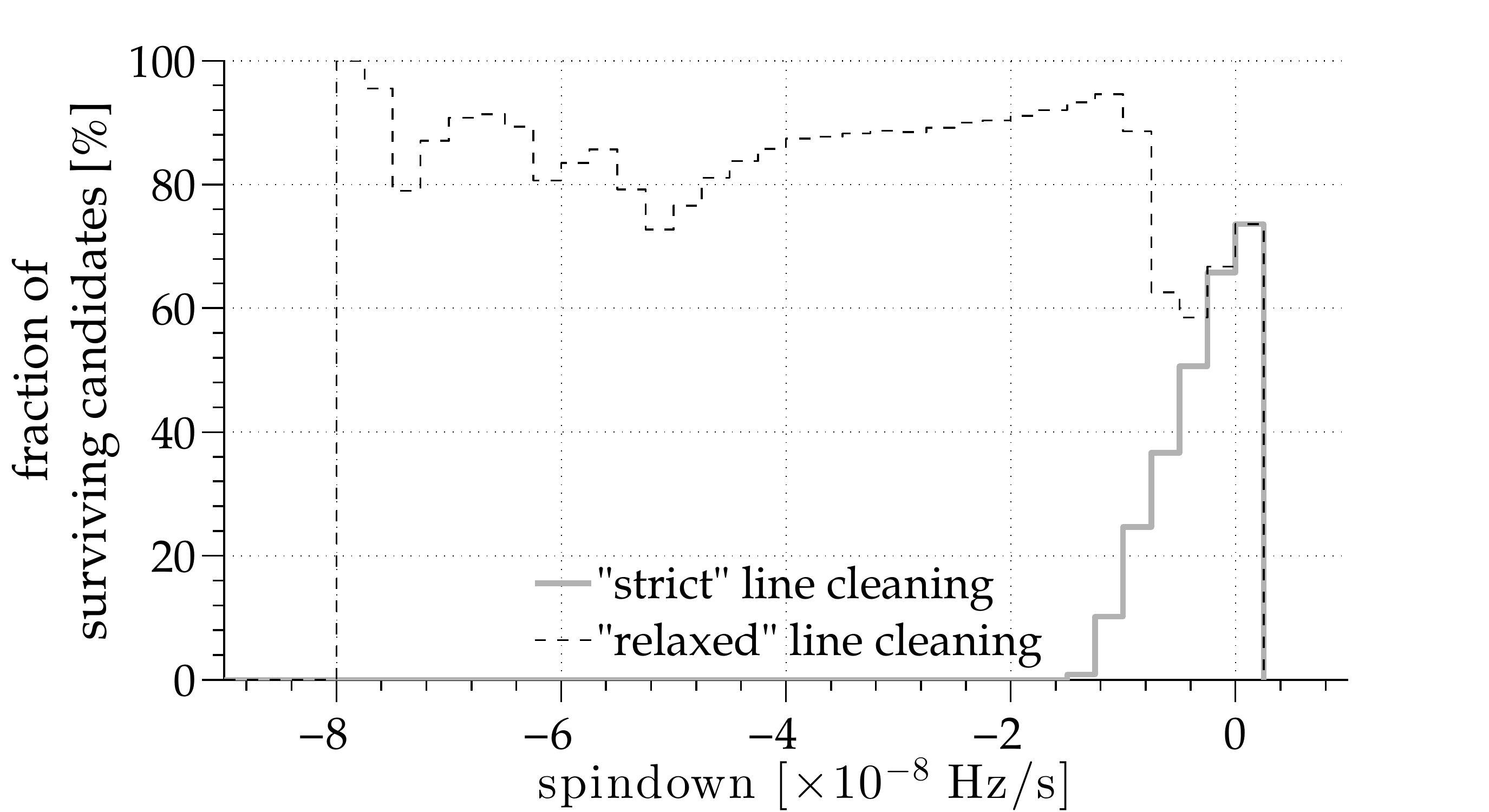}
	\caption{The histograms illustrate the effect of the relaxed known-lines cleaning
procedure in
the context of the \GCSearch{}: the black, dashed histogram contains the fraction of candidates that
pass
the strict known-lines cleaning, the gray, solid histogram shows the fraction of candidates that
pass
the relaxed
known-lines cleaning. }
\label{fig:cleaningcomparison}
\end{figure}

After applying the relaxed known-lines cleaning in the \GCSearch{}, $83\%$ of the candidates
survive.
Figure~\ref{fig:cleaningcomparison} shows the fraction of surviving candidates as a function of
their
spin-down values for both the strict and the relaxed known-lines cleaning procedure.

%
%
\subsection{Clustering of candidates}
\label{lab:clustering}

A GW signal that produces a significant value of the detection statistic at the template with the
closest parameter match will also produce elevated detection statistic values at templates
neighboring
the best-match template. This will also happen for a large class of noise artifacts in the
data. The clustering procedure groups together candidates around a local maximum, ascribing them to
the same physical cause and diminishing the multiplicity of candidates to be considered in the
following steps. The clustering is performed based on the putative properties of the signals:
assuming a
local maximum of the detection statistic due to a signal, the cluster would contain with high
probability all the parameter-space points around the peak with values of the detection statistic
within at least half of the peak value.

\begin{figure}[]
	\includegraphics[trim= 5mm 0mm 5mm 0mm, clip,
width=0.5\textwidth]{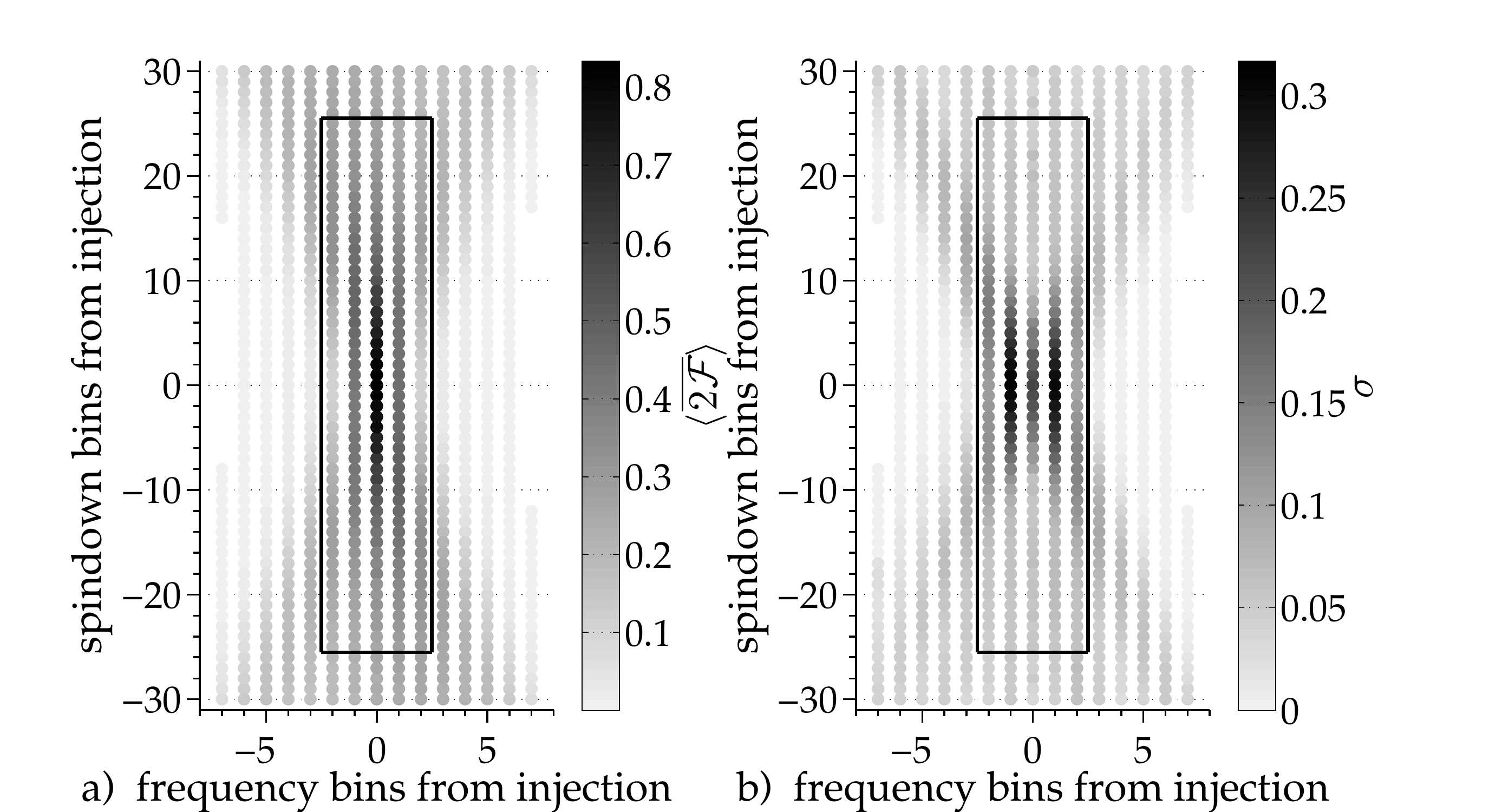}
	\caption{The left plot shows the averaged outcome of 200 searches that were performed on fake
data with parameters randomly distributed over the search parameter space. As explained in the text,
the
detection statistic values are all normalized to the maximum of their respective search and then
averaged over the
200 searches in each grid point. The grid points represent distances from the maximum. The right
plot gives the standard deviation for each data point of the left plot. The black box denotes the
cluster box. All candidates within this box are ascribed to the loudest representative candidate in
the center by the clustering procedure.}
	\label{fig:averageandfluctuation}
\end{figure}

The cluster is constructed as a rectangle in frequency and spin-down. We choose to use the bin size
of Eq.~\ref{eq:grid-spacings} as a measure for the cluster box dimension. The cluster box size
is determined by Monte Carlo simulations in which 200 different realizations of CW
signals randomly distributed within the ranges given in Table \ref{tab:falsedismissalparams} and
without
noise are created and the data
searched. The search grid is randomly mismatched with
respect
to the signal parameters for every injection. Figure~\ref{fig:averageandfluctuation}a shows the
average detection
statistic values, each normalized to the value of the maximum for each search, as a function of the
frequency and spin-down distance from the location of the maximum. The average is over the 200
searches. Figure~\ref{fig:averageandfluctuation}b shows the standard deviation of the normalized
detection statistic averaged in order to determine Fig.~\ref{fig:averageandfluctuation}a.

\begin{figure}[]
	\includegraphics[trim= 5mm 0mm 0mm 0mm, clip, width=0.5\textwidth]{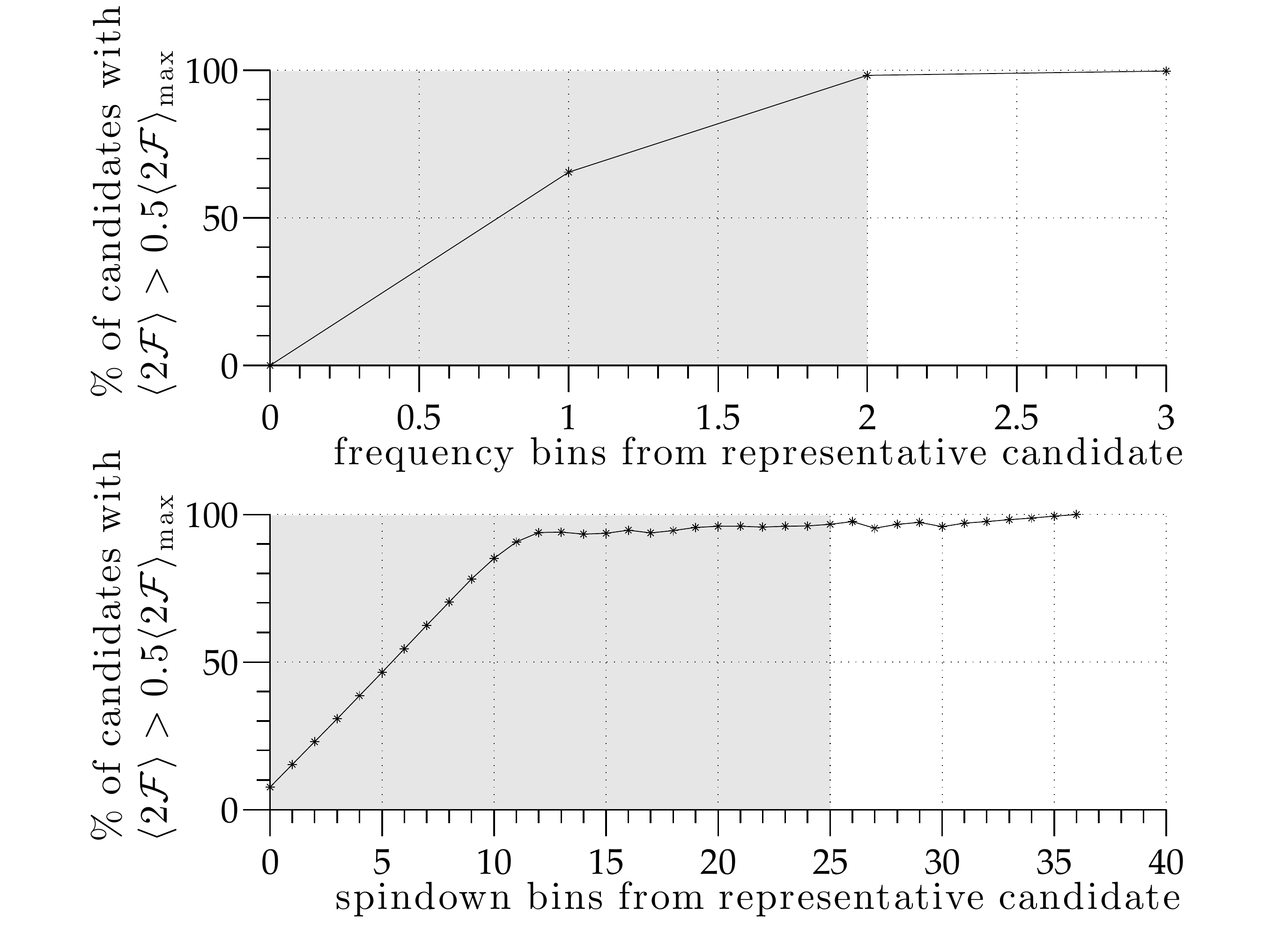}
	\caption{The plot shows the fraction of candidates with $\satf$ values larger than half of
maximum as a function of distance in frequency (top) and spin-down (bottom), from the maximum. The
gray shaded areas denote the
boundaries of the cluster box in each dimension.}
	\label{fig:averageclustersize}
\end{figure}

Figure~\ref{fig:averageclustersize} shows the distribution of the distances in frequency and
spin-down bins for average normalized detection statistic values greater than 0.5. Within the
neighboring two frequency bins on either sides of the frequency of the maximum, we find $>99\%$ of
the templates with average normalized detection statistic values larger than half of the maximum. Of
order $95\%$ of the templates that lie within 25 spin-down bins of the spin-down value of the
maximum have average normalized detection statistic values larger than half of the maximum.
Therefore, we pick the cluster size to be
\begin{align}  \label{eq:clusterbox}
	&\Delta f_{\text{cluster}} = \pm 2\ \delta f \ \simeq\  4.8\times 10^{-5}\,\Hz\nonumber\\
	&\Delta\dot f_{\text{cluster}} = \pm 25\ \delta \dot f_\text{fine} \ \simeq\ 4.5\times
10^{-12}\,\Hz/\s
\end{align}
on either side of the parameter values of the representative candidate.

The clustering procedure is implemented as follows: from the list of all candidates surviving the
known-lines cleaning the candidate with the largest $\satf$ value is chosen to be the representative
for the first cluster. All candidates that lie within the cluster size in frequency and spin-down
are
associated with this cluster. Those candidates are removed from the list. Among the remaining
candidates the one with the largest $\satf$ value is identified and taken as the representative of
the second cluster. Again, all candidates within the cluster size are associated to that cluster and
removed from the list. This procedure is repeated until all candidates have either been chosen as
representatives or are associated to a cluster.

After applying the clustering procedure to the list of candidates in the context of
the \GCSearch{}, the total number of candidates left for further postprocessing checks is reduced
to \mbox{$\sim 33\%$} of what it was before the clustering procedure.

%
\subsection{The $\F$-statistic consistency veto}
\label{lab:consistency}

The $\F$-statistic consistency veto provides a powerful test of local disturbances by testing
for consistency between the
multidetector detection-statistic results and the single-detector results. Local disturbances are
more likely to affect the
data of only one detector and appear more clearly in the single-detector results than in the
multidetector results. In contrast, GWs
will be better visible when the combined data set is used. The veto which has already been used in
past CW searches
\cite{Aasi:2012fw,Abbott2009d} is simple: for each candidate the single-interferometer
values $\satf_\text{H1,L1}$ are computed. The results are compared with the
multi-in\-ter\-fe\-ro\-me\-ter $\satf$ result. If any of the single-in\-ter\-fe\-ro\-me\-ter values
is higher than the multi-in\-ter\-fe\-ro\-me\-ter value, we conclude that its origin is local and
the associated candidate is discarded.

While designing the different steps to be as effective as possible in rejecting disturbances, it is
important to make sure that a real GW signal would pass the applied vetoes with high probability. To
this end the
vetoes are tested on a set of simulated data with signal injections and the false-dismissal rate is
estimated. In
particular, we create $500$ different data sets consisting of Gaussian noise
and signals with parameters randomly distributed within
the search parameter space of the \GCSearch{}. An overview of the
injection parameter ranges is given in Table~\ref{tab:falsedismissalparams}. The SFT start times of
the test
data set equal those of the real data set. Unless stated otherwise,
the same procedure is used whenever injection studies are reported throughout the paper. The data
are searched with the same template grid as used in the original search, in a
small region around the signal parameters, and the maximum from that search is taken as the
resulting
candidate.

\begin{table}[tbp]
    \centering
  \begin{tabular*}{0.48\textwidth}{rcl}
	\hline
    Parameter & \phantom{.} & Range \\
	\hline
    Signal strength & & $h_0^\text{injected} = h_0^{90\%}(f)$ of the \GCSearch{}\\
	Sky position [rad] & & $\lvert\Delta\alpha\rvert + \lvert\Delta\delta\rvert \leq
10^{-3}\,\text{rad}$ \\
	&& with $\Delta\{\alpha,\delta\} = \{\alpha,\delta\}_\text{inj} - \{\alpha,\delta\}_\text{GC}$\\
	Frequency [Hz] & & $150\,\Hz \le f \le 152\,\Hz$\\
	Spin-down [Hz/s] & & $-f/200\ \yr \leq \dot f\leq 0$\\
	Polarization angle & & $0\le \psi \le 2\pi$\\
	Initial phase constant & & $0\le \phi_0 \le 2\pi$\\
	Inclination angle & & $-1\le \cos\iota \le 1$\\
    \hline
  \end{tabular*}
  \caption{Parameters of the fake signal injections performed for the false-dismissal studies.
Unless explicitly stated in the text, $500$ signals are
injected into fake Gaussian noise.}
  \label{tab:falsedismissalparams}
\end{table}

The $\F$-statistic consistency veto is found to be very safe against false-dismissals: none of the
$500$ signal
injections is vetoed, which implies a false-dismissal rate of $\lesssim 0.2\%$ for the
tested population of signals.

In the \GCSearch{}, this veto removes $12\%$ of the tested candidates.

%
\subsection{The significance threshold}
\label{lab:significanceThreshold}

At some point in most CW searches, a significance threshold is established below which candidates
are
not considered. This threshold limits the ultimate sensitivity of the search. Where one sets this
threshold depends on the available resources for veto and follow-up studies. If no resources are
available beyond the ones used for the main search, this detection threshold will be based solely on
the results from that first search.
If there are computing resources available to devote to follow-up studies, the threshold should be
the lowest
such that candidates at that threshold level would not be falsely discarded by the next stage and
such that
the follow-up of all resulting candidates is feasible. From these considerations follows a
significance threshold of $\satf \ge 4.77$ for the \GCSearch{} \cite{myobspaper}. This reduces the
number of candidates to follow up to a number of 27\,607 which, after the next veto and excluding
the candidates
that we can associate to fake signals present in the data stream for validation purposes, becomes
manageable for a
sensitive enough follow-up.
In the absence of surviving candidates at any threshold one can always set upper limits based on the
loudest surviving candidate below the threshold, assuming that it is due to a putative
signal.

%
\subsection{The permanence veto}
\label{lab:permanence}

Until now we have not required the searched GW signals to have any specific characteristics other
than having consistent properties among the two detectors used for the analysis.
To further reduce the number of candidates to follow-up (and hence attain a higher sensitivity) we
now restrict our search
to strictly continuous signals and use their assumed permanence during the observation
time to define the next veto. We stress that GW
signals such as  strong transient GW signals \cite{prixetal2011:_transientCW,2011PhRvD..83h3004T}
lasting a few days or
weeks which might have survived
up to this point
would be dismissed by this veto. With this step we trade breadth of search in favor of depth.

In a stack-slide search the detection statistic $\satf$ is the average of the $2\F$ values
computed for each of the segments and we expect that the SNR of a signal to be
comparable in every segment.
However, in
the final result the relative contribution of the different segments to this average  is
``invisible.''
This information is important, though, because a strong-enough disturbance can cause a large
$\satf$ value, even though only a few segments effectively contribute to it. The following veto aims
to uncover behavior of this type and discard the associated candidates. \par

In the context of the \GCSearch{} we found that in most cases it is only a single segment which
contributes a
large $\satf$ value.
This observation inspired the definition of the simple veto that was used: for each candidate the
values of $2\F$ for each of the $630$ data segments are determined. The highest value is removed and
a new average over the remaining $629$ segments is computed. If this value is below the significance
threshold (see Sec.~\ref{lab:significanceThreshold}), the candidate is vetoed.

\begin{figure}[]
	\includegraphics[trim= 15mm 0mm 0mm 0mm, clip, width=0.5\textwidth]{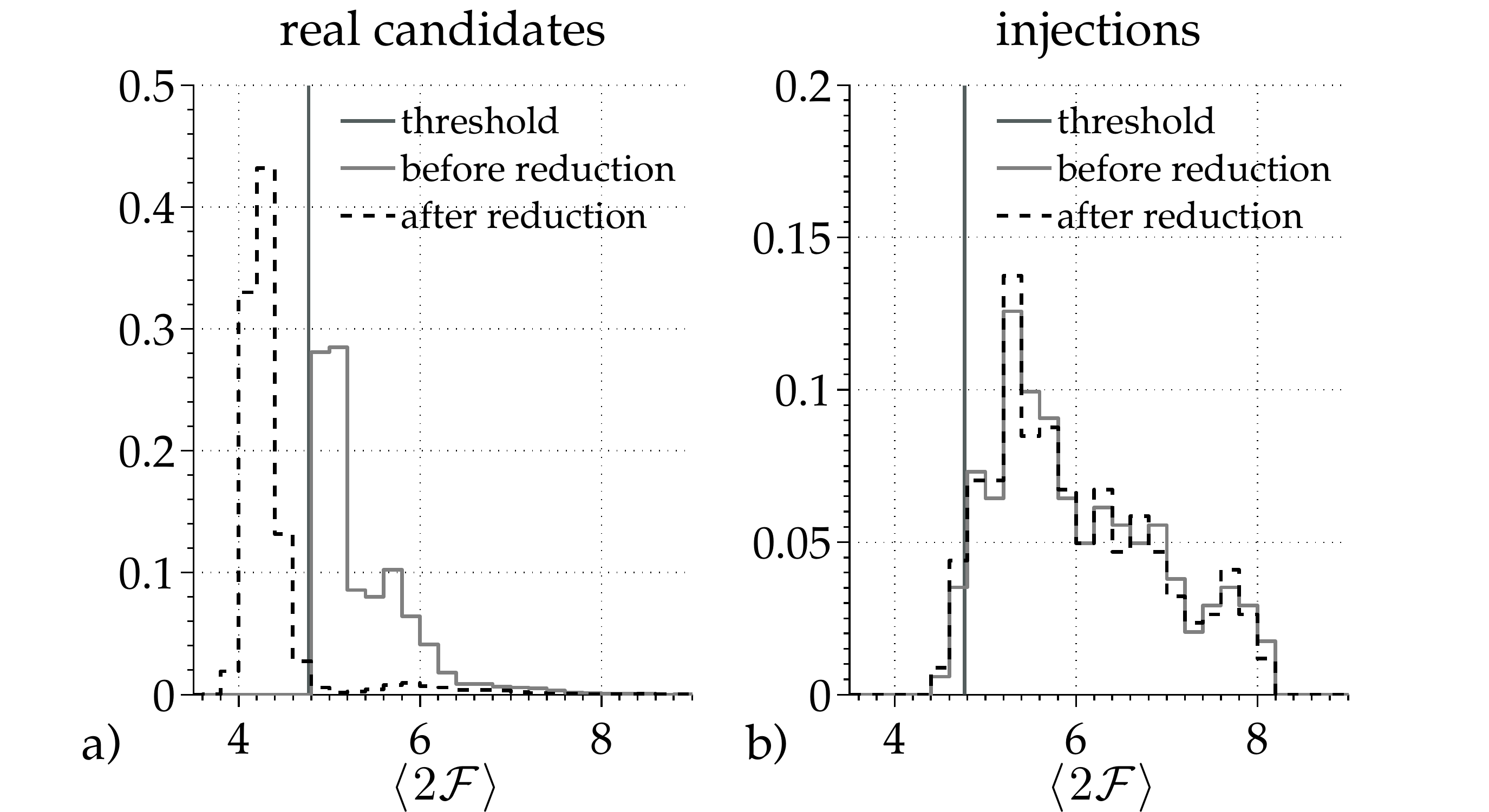}
	\caption{The distribution of $\satf$ values before and after removing the contribution from the
	loudest among the different segments (a) for real
candidates and (b) for a set of signal injections. Removing the contribution of the loudest
segment shifts the
distribution of $\satf$ values from just above to below the significance threshold at $\satf=4.77$
(vertical line). This
clearly shows that most candidates have significant $\satf$ values only due to a single segment that
contributes an extremely large $2\F$ value. The distribution of $\satf$ values from a set of
continuous signal
injections does not change: the contribution of all segments to the average value is comparable.}
	\label{fig:permanencehistograms}
\end{figure}

The permanence veto was highly effective in the \GCSearch{} where it ruled out about 96\% of the
candidates from the previous stage. At the same time it is very safe, with a false-dismissal rate of
only $\simeq 0.6\%$ for the given signal population (see Table \ref{tab:falsedismissalparams}).
Figure~\ref{fig:permanencehistograms} shows the distribution of the $\satf$ values before and after
removing the loudest segment contribution. One can clearly see that removing the loudest
contribution shifts the
distribution from above to below the threshold for most of the candidates that were tested with this
veto in
the context of the \GCSearch{} (\ref{fig:permanencehistograms}a), but does not change the
distribution when applied to a set of signal injections (\ref{fig:permanencehistograms}b).

%
%
\subsection{Coherent follow-up search and veto}
\label{lab:followup}

The semicoherent technique used in the \GCSearch{} is a powerful way to search a large parameter
space, but this benefit comes at the cost of reduced sensitivity: a semicoherent analysis does
not recover weak signals with the same significance as a coherent search technique could do on the
same data set
and with comparable mismatch distributions, and it does not estimate the parameters of a signal as
precisely as a coherent analysis would.
We recall that the advantage of using semicoherent techniques is that they allow the probing of
large
parameter-space volumes, over large data sets
with realistic amounts of computing power.

\begin{figure}
	\includegraphics[trim = 15mm 0mm 0mm 0mm, clip, width=0.5\textwidth]{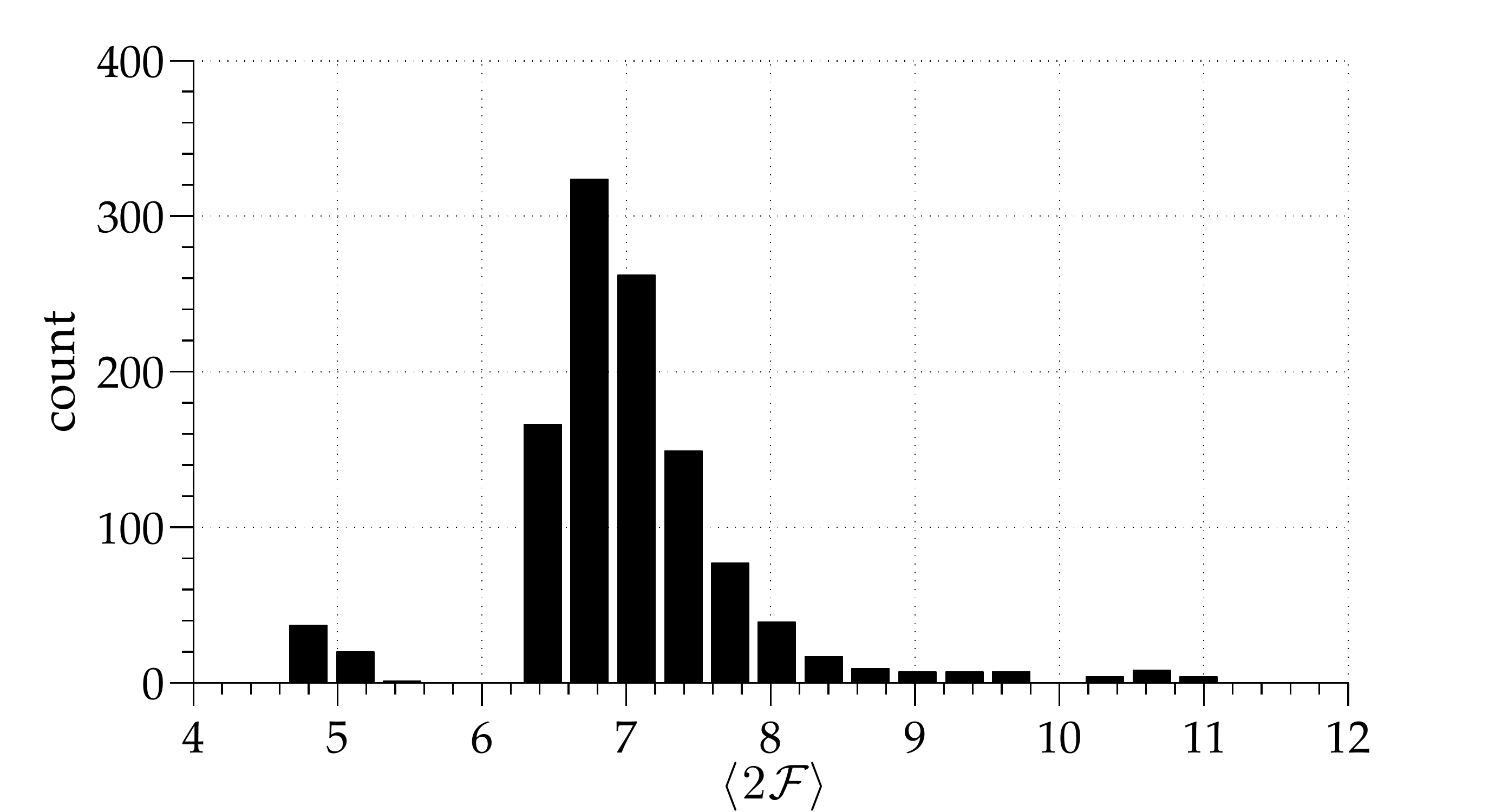}
	\caption{The histogram shows the distribution of the $\satf$ values of the 1\,138 candidates
that
are investigated in the follow-up search. All candidates with values larger than 6 can be ascribed
to a pulsar signal hardware injection that was performed during S5.}
	\label{fig:followupcandshist}
\end{figure}

A coherent follow-up search can now be used on a subset of the original parameter space identified
as
significant by the previous stages (see Fig.~\ref{fig:followupcandshist}). By appropriately choosing
the coherent observation time, we ensure that this search rules
out with high confidence candidate signals that fall short of a projected detection statistic value.

The process of determining the optimal coherent observation time, under constraints on computing
power and on the precision of the candidate signals' parameters, is iterative. In the \GCSearch{} a
setup was
determined that was able to coherently follow up the remaining 1\,138 candidates, having allocated
on
average 10\,h of computing time for each
candidate. The chosen data set spanned $T\coh=90$~days and yielded a moderate computing cost of the
order of a day on several hundred compute nodes of the ATLAS cluster.

In the following we detail a single-stage follow-up procedure with this coherent observation time,
illustrate the condition used to test the candidates after the follow-up and demonstrate its
effectiveness.
\begin{figure}
	\includegraphics[trim = 10mm 0mm 10mm 0mm, clip, width=0.5\textwidth]{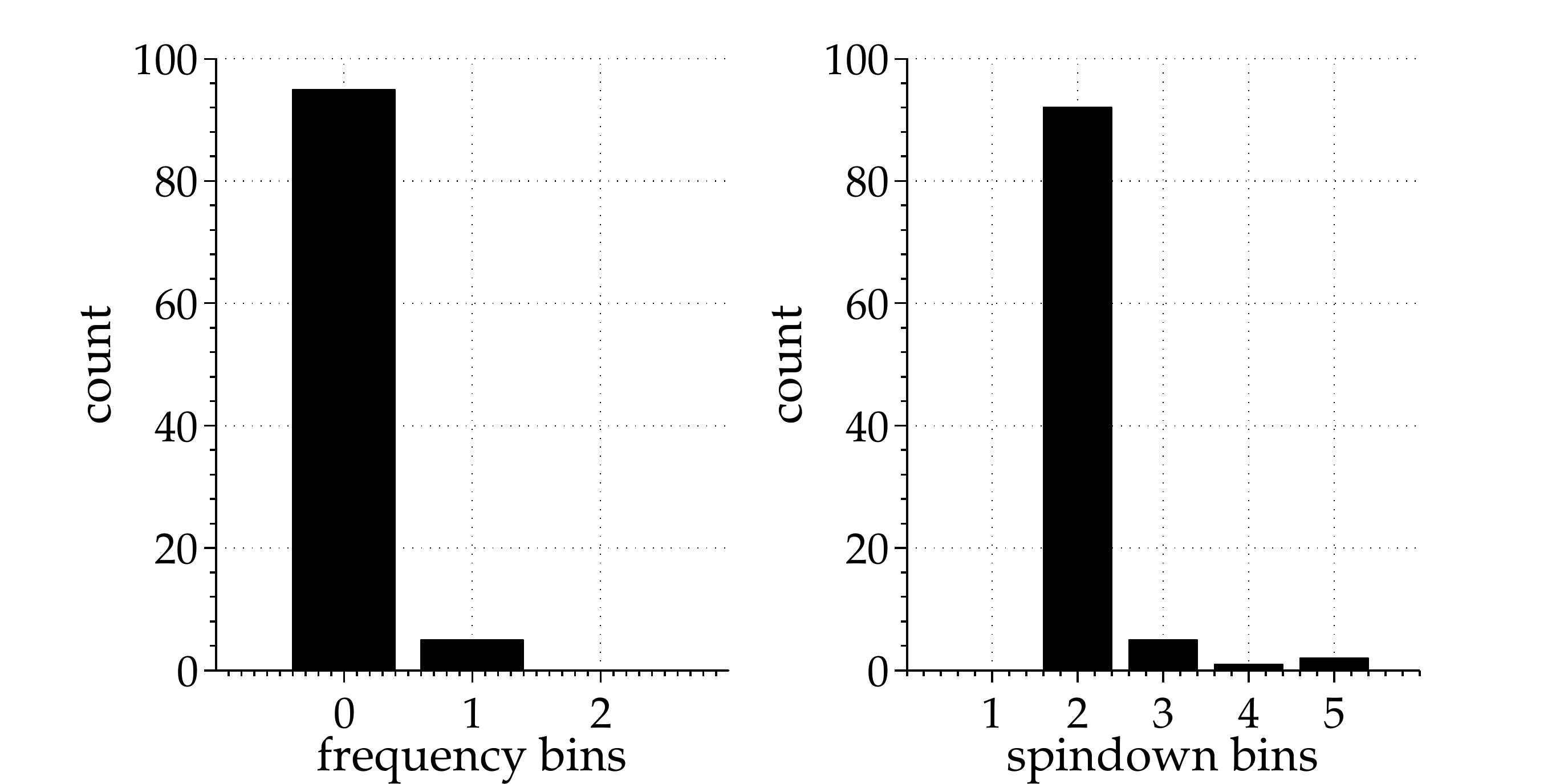}
	\caption{The two histograms show the distances (absolute value) between the parameters of the
loudest template
and the parameters of the injected signal measured  in frequency and spin-down bins of the initial
search [i.e. with the spacing of Eq.~\ref{eq:grid-spacings}]. The counts refer to values of the
distances between the right and left edges of each bin.}
	\label{fig:boxsize}
\end{figure}

\subsubsection{The search volume}
\label{sec:followup_searchbox}

The parameters of the candidates are only approximations to the real signal's parameters. In order
to establish the parameter-space region that we need to search around each candidate in the
follow-up search, we have studied the distribution of distances between the injection parameters and
the recovered parameters for a population of fake signals designed to simulate the outcome of the
original search.

In particular, we have performed a Monte Carlo study over $100$ fake signals with parameters
distributed as described in Table~\ref{tab:falsedismissalparams} without noise.
The data are searched with the template grid resolution of the \GCSearch{} in a box of
$50$ frequency bins and $100$ spin-down bins, placed around each injection. Then the distance in
frequency and spin-down between the loudest candidate and the injection is evaluated.
The distributions of these distances are shown in the histograms in Fig.~\ref{fig:boxsize}. In all
cases, the distance is smaller or equal to two frequency and five spin-down bins. Consequently, the
frequency and spin-down ranges for the coherent follow-up search are set to be:
\begin{align}
\label{eq:followupbox}
	&\Delta f = \pm 2\ \delta f \rightarrow 1.2 \times 10^{-4}\,\Hz\nonumber\\
	&\Delta\dot f = \pm 5\ \delta \dot f_\text{fine}  \rightarrow 2.0  \times 10^{-12}\,\Hz/\s.
\end{align}

In order to keep the computational cost of the follow-up stage within the set bounds we restrict the
search sky region to a distance of 3\,pc around the
GC.

\subsubsection{The grid spacings}
\label{sec:followup_grid}

The frequency-spin-down region in parameter space around each candidate defined by
Eq.~\ref{eq:followupbox} is covered by a template grid refined with respect to the original one of
Eq.~\ref{eq:grid-spacings}, as follows:
\begin{align}
\label{eq:followupspacings}
	&\delta f_\text{coh} = (2\ T_{\text{coh}})^{-1} = 6.4 \times 10^{-8}\,\Hz\nonumber\\
	&\delta \dot f_\text{coh} = (2\ T_{\text{coh}}^2)^{-1}= 8.3 \times 10^{-15}\,\Hz/\s.
\end{align}

The sky region to be searched is covered by a $6\,\times \,6$ rectangular grid in right ascension
and
declination, across $7.2 \times 10^{-4}$ rad in each dimension.

\begin{figure}[]
	\includegraphics[trim = 0mm 0mm 0mm 0mm, clip, width=0.5\textwidth]{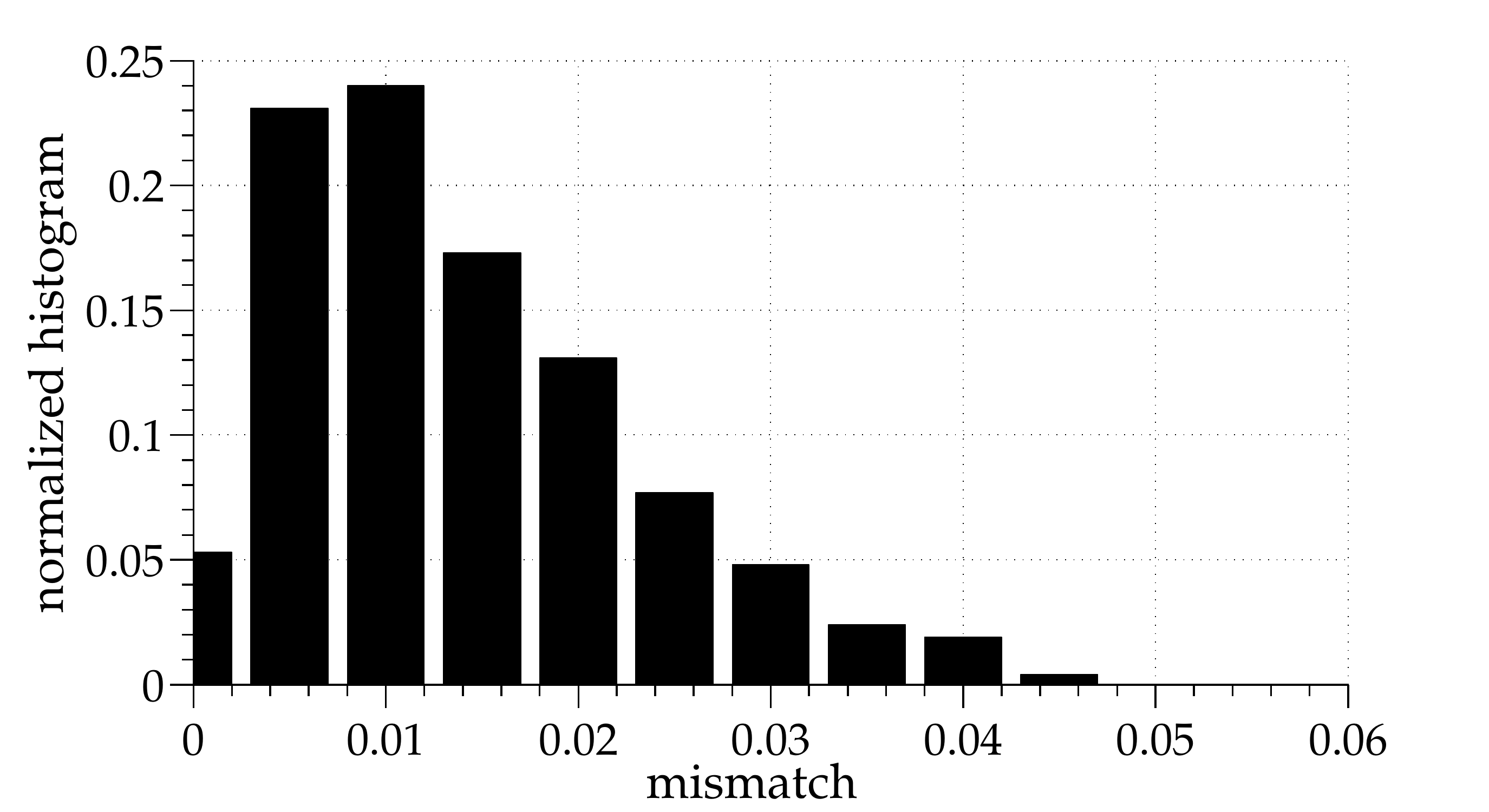}
	\caption{This histogram shows the mismatch distribution of the follow-up search. The average
value is $\langle m_\text{coh}\rangle = 1\%$. Only in a small fraction of cases ($\lesssim 1\%$) can
the
mismatch be as high as 4.7\%.}
	\label{fig:followupmismatch}
\end{figure}

With all these choices the resulting mismatch distribution over 1\,000 trials in Gaussian noise is
shown in
Fig.~\ref{fig:followupmismatch}. It has an average of $\langle m_\text{coh}\rangle = 1\%$ and
reaches 4.7\% in $\lesssim 1\%$ of the cases.

\subsubsection{The expected highest $2\F$ in Gaussian noise}
\label{sec:followup_loudestinnoise}

\begin{figure}[]
	\includegraphics[trim = 0mm 0mm 0mm 0mm, clip,
width=0.5\textwidth]{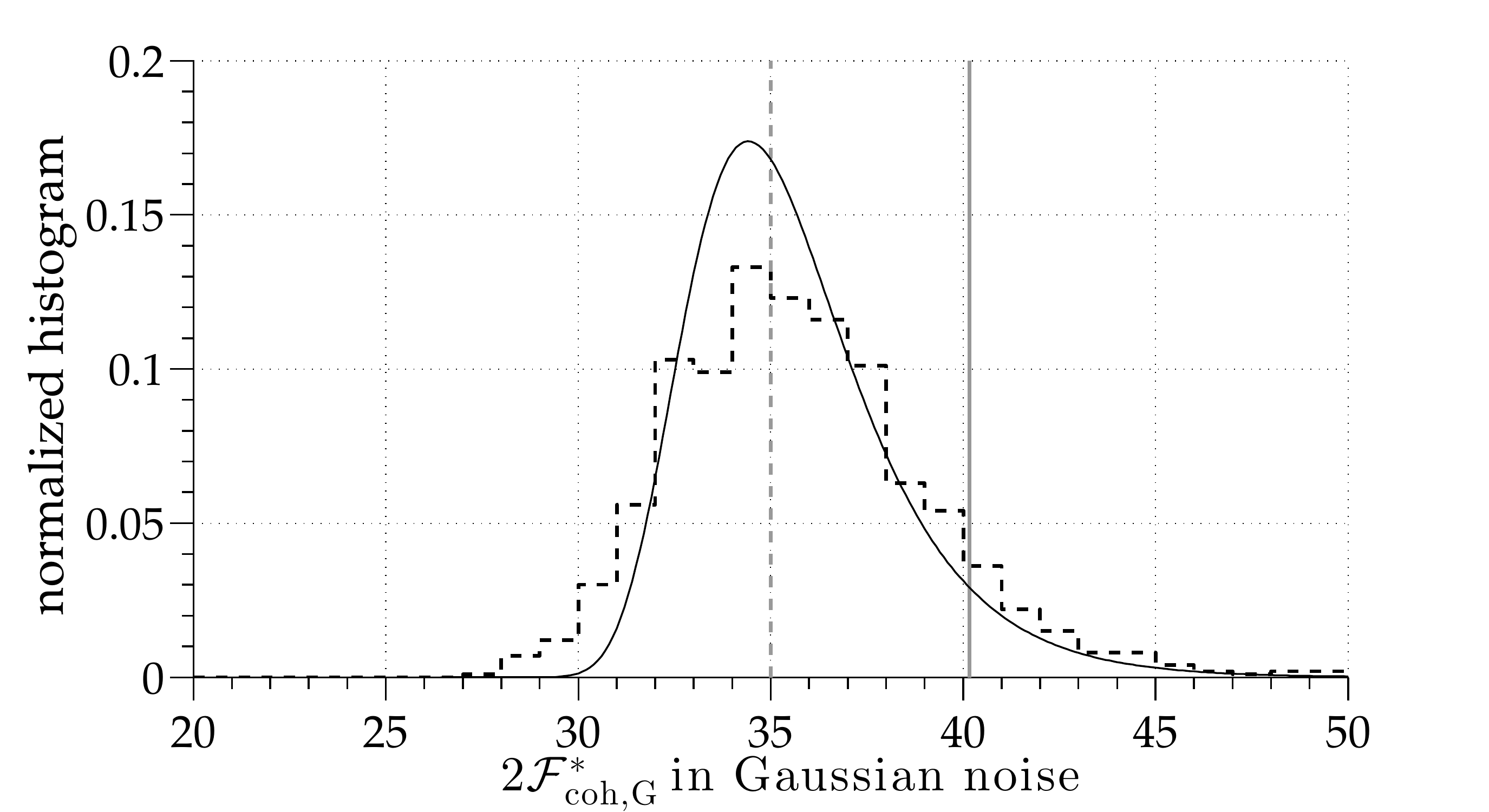}
	\caption{This histogram shows the distribution of the maximum $2\F^{\ast}_\text{coh,G}$
resulting from a coherent follow-up search. A thousand searches with grids defined by
Eq.~\ref{eq:followupspacings} were performed, with different realizations of pure Gaussian noise
(black dashed line). The mean value measured on this distribution is
$35\pm3$ (dashed gray line). Based on the total number of templates $N_\text{coh}$, the predicted
mean is $E[2\F^\ast_\text{coh, G}] = 40\pm4$ (solid gray line).
From the simulation data the number of effective independent templates can be
estimated to $N_\text{eff}\sim 0.1\ N_\text{coh}$ (black line). The solid black line is the
estimated probability density function for the maximum under the assumption of $N_\text{eff}$
independent templates.}
	\label{fig:followupgaussiannoise}
\end{figure}

The follow-up of the region of parameter space around each candidate defined in
Eq.~\ref{eq:followupbox} and covered by the grids defined in Eq.~\ref{eq:followupspacings} results
in a total of $N_\text{coh} = N_f \times N_{\dot f} \times N_\text{sky} \sim 16 \times 10^{6}$
templates. The expected largest $2\F$ value for Gaussian noise and for $N_\text{coh}$ independent
trials is $E[2\F^{\ast}_\text{coh, G}] = 40$ \cite{myobspaper} with a standard deviation
$\sigma_G=4$.
However, this is
an overestimate since the templates in our grids are not all independent of one another. We estimate
the {\it effective} number of independent templates through the actual measured distribution. This
is obtained by a Monte Carlo study in which $1\,000$ different realizations of Gaussian noise are
analyzed, using the search box and the template setup of the coherent follow-up search. The loudest
detection statistic value from each search is recorded. Figure~\ref{fig:followupgaussiannoise} shows
the distribution of such largest $2\F^{\ast}_\text{coh,G}$ values. The measured mean value is $35$
and the measured standard deviation is $3$. We superimpose the analytic expression for
the expectation value [see \cite{Wette2010}, Eq.~7] for an effective number of templates
$N_\text{eff} \sim 0.1\ N_\text{coh}$. The analytic estimate with $N_\text{eff}$ has a mean
value of 36 and a standard deviation of 3.

\subsubsection{The expected detection statistic for signals}
\label{sec:followup_expected2F}

In the presence of a signal the coherent detection statistic $2\F$ follows a $\chi^2$ distribution
with four degrees of freedom and a noncentrality parameter $\rho^2$ which depends on the noise
floor and on the signal amplitude parameters and scales linearly with the coherent observation time.
The expected value of $2\F$ is $\mathrm{E}[2\F] = 4 + \rho^2$.

For every candidate the $\satf$ value can be expressed in terms of the
$\langle\rho^2_{\text{cand}}\rangle$ associated with that
candidate:
$\langle\rho^2_{\text{cand}}\rangle=\satf_{\text{cand}}-4$. Based on
$\langle\rho^2_{\text{cand}}\rangle$ we
then estimate the expected detection statistic of a coherent follow-up with $T_\text{coh}$ data
simply as\footnote{This is valid for data sets with equal fill level, as is the case in
the \GCSearch{}. For other data sets, instead of $T_\text{coh}/T_\seg$ the actual amount of data
needs to be taken, $\left( N_\seg T_\text{coh}^\text{data} \right)/ T_\text{orig}^\text{data}$. In
Eq. \ref{eq:expvalueequationnew} we have also
neglected the ratio of the noise ${S_h^{\text{j,orig}}} / {S_h^{\text{i,coh}}}$, based on our
observations on the specific data set used.}
\be
\label{eq:expvalueequationnew}
	2\F^{\text{cand}}_\text{coh} = \kappa\left( \langle\rho^2\cand
\rangle\frac{T_\text{coh}}{T_\text{seg}}
\right) +
4,
\ee
where $\kappa$ is the ratio between the sum over the antenna pattern functions for each SFT $i$ used
by the
coherent follow-up search and the SFTs $j$ of the original search, respectively:
\be
\label{eq:computekappa}
\kappa= \frac{\sum_i \left( F_+^2 + F_\times^2 \right)_{i,\text{coh}}}{\sum_j \left( F_+^2 +
F_\times^2 \right)_{j,\text{orig}}}.
\ee
This antenna pattern correction $\kappa$ is important, because it accounts for intrinsic differences
in sensitivity to a source at the GC between the data set of the original search and the data set
used for the follow-up. For the original search, a data-selection procedure was used to construct
the
segments at times when the detectors were particularly sensitive to the GC. In contrast,
$T_\text{coh}$ spans many weeks and also comprises data that were recorded when the orientation
between the detector and the GC was less favorable.
For this reason, a simple extrapolation of the $\satf\cand$ values from the original search to
estimate the result of the follow-up search without such a correction, folding in implicitly the
assumption that the antenna pattern values for the data of the original search data set are
equivalent to those of the
follow-up
search data set, would result in significantly wrong predictions.

The chosen data set of the follow-up search comprised data from the H1 and L1 detectors and spanned
90
days in the time between February~1, 2007, at 15:02~GMT and May~2, 2007, at 15:02~GMT. It
contained
a total of $6\,522$~SFTs ($3\,489$ from H1 and $3\,033$ from L1) which is an average of $67.9$~days
from
each detector\footnote{The data are chosen by the same procedure as described in
Sec.~\ref{lab:dataselection}, but this time by grouping the SFTs into
segments of $90$~days. Neighboring segments overlap each other by $24$\,h.} (fill level of 75.5\%).
The antenna pattern correction for this data set and the one used in the original search can be
computed with Eq.~\ref{eq:computekappa} and results in $\kappa = 0.68$.
The fact that $\kappa < 1$ shows the effect of having chosen data segments for the original search
that
were recorded at times with favorable orientation between the detectors and the GC, whereas in a
contiguous
data stretch as long as 90 days the antenna patterns average out.

\subsubsection{The $\R$ veto}

\begin{figure}
	\includegraphics[trim = 0mm 0mm 10mm 0mm,
clip,width=0.5\textwidth]{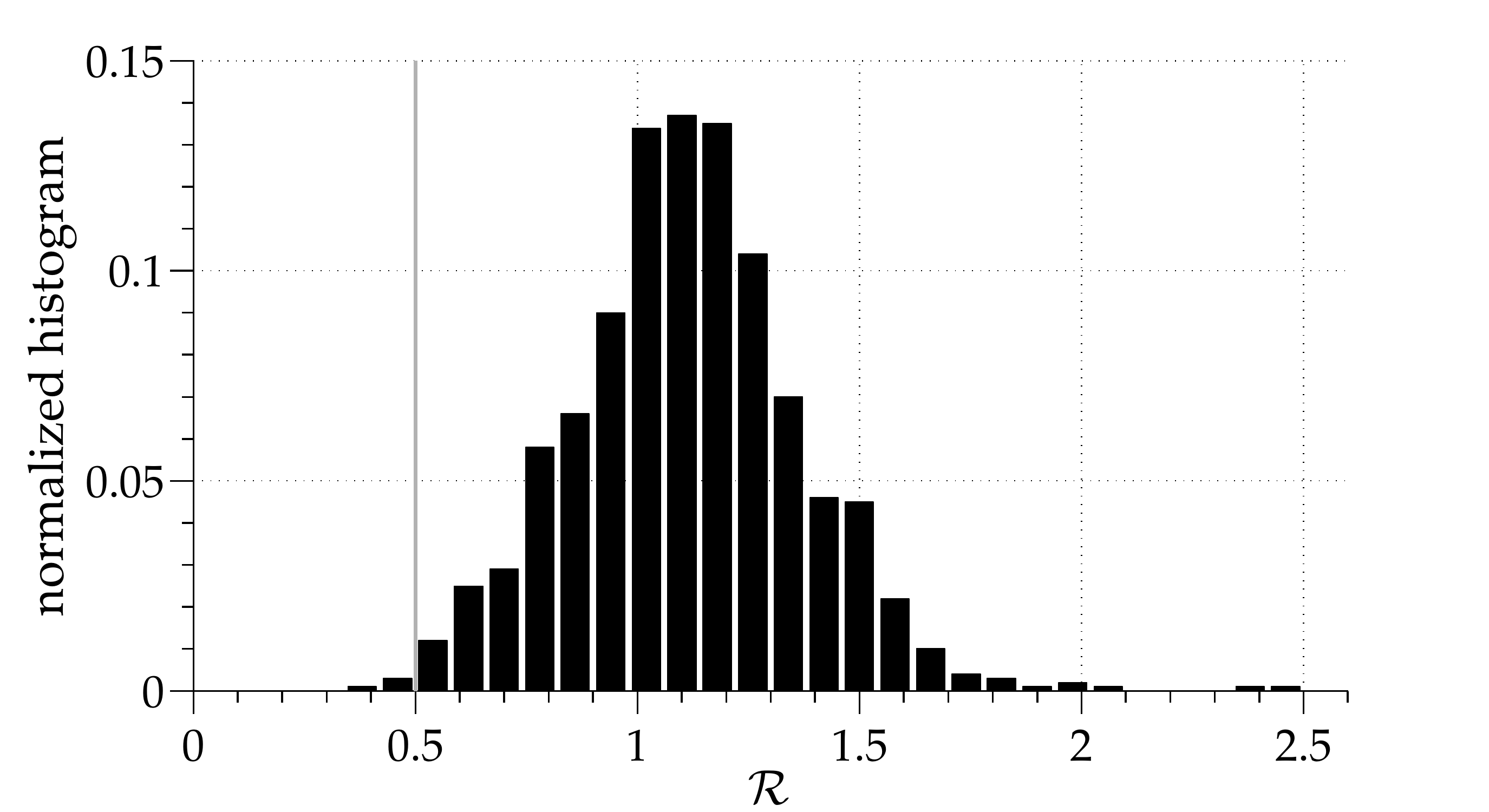}
	\caption{False dismissal: The plot shows the distribution of $\R$ for $1\,000$ different
realizations of CW signals in Gaussian noise with parameters randomly distributed within the search
parameter space. The threshold is set to $\mathcal{R}_\text{thr} = 0.5$ (gray line). Only four of
the
$1\,000$ signals have $\mathcal{R} < \mathcal{R}_\text{thr}$, implying a false-dismissal rate of
$0.4\%$.}
	\label{fig:falsedismissalfollowup}
\end{figure}

The expected $2\F^\text{cand}_{\text{coh}}$ for each candidate is computed. A follow-up search is
performed as specified in the previous sections and the resulting highest value
$2\F^\ast_\text{coh}$
is identified. We define the ratio
\be\label{eq:ratio}
\R := \frac{2\F^\ast_\text{coh}}{2\F^\text{cand}_\text{coh}}.
\ee
A threshold $\R_\text{thr}$ is set on $\R$ and candidates with $\R < \R_\text{thr}$ are ruled
out as GW candidates: their measured detection statistic value after the follow-up falls short of
the predictions. The threshold is obtained empirically with a Monte Carlo study with $1\,000$ signal
injections in Gaussian noise. The signal parameters are uniformly randomly
distributed within the ranges given in Table~\ref{tab:falsedismissalparams}. Two separate data sets
are created for this study: one that matches the original data set (in terms of SFT start times)
and one that matches the $90$-days coherent data set.
Figure~\ref{fig:falsedismissalfollowup} shows the distribution of the corresponding values of $\R$.
The peak of the distribution of $\mathcal{R}$ is slightly above 1. This means that we slightly
underestimate
the outcome of the coherent follow-up search.

We place the threshold at $\R_\text{thr} = 0.5$. The plot shows that $\R < 0.5$ for only four out of
the $1\,000$ injected signals. This implies a false-dismissal rate of $\sim0.4\%$.

\begin{figure}
	\includegraphics[trim = 0mm 0mm 10mm 0mm, clip,width=0.5\textwidth]{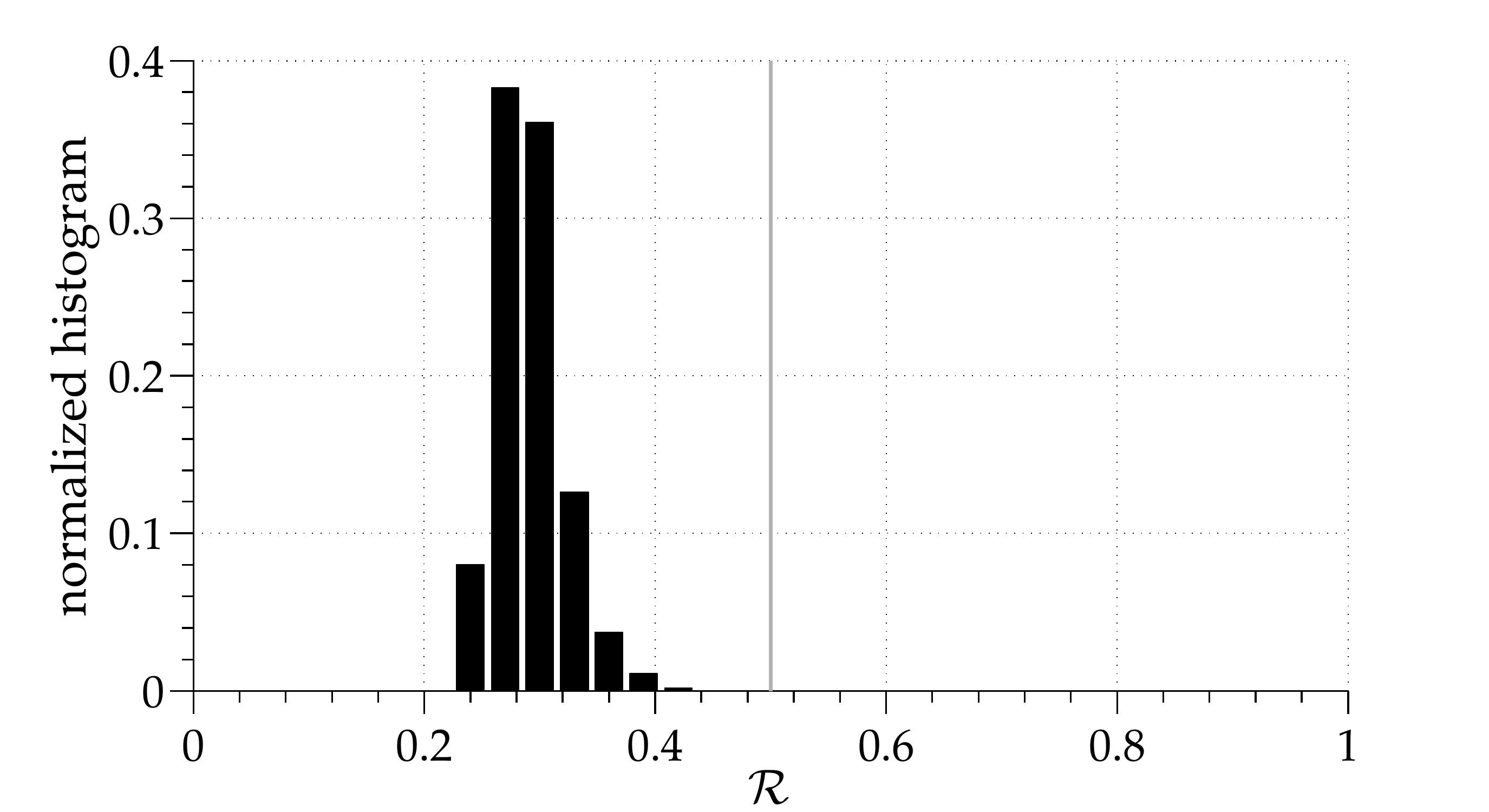}
	\caption{False alarm: The least significant candidates that are considered in the follow-up have
$2\F$ values at the significance threshold. Those candidates are expected to result in a value of
$2\F^{0.77}_\text{coh} = 105.83$ in the follow-up search if they are due to a CW. This plot shows
the distribution of $\mathcal{R}$ values resulting from a follow-up search on 1\,000 different
realizations of Gaussian noise.
None of the candidates satisfied the condition $2\F^\ast_\text{coh}  > 52.9$, which implies a
Gaussian false-alarm rate of $\leq 0.1\%$.}
	\label{fig:falsealarm}
\end{figure}

Although a candidate that would pass the coherent follow-up $\mathcal{R}$ veto would still need to
undergo further checks to be confirmed as a GW signal, it would all the same be exciting. To
estimate the chance of a false-alarm event in Gaussian noise, a Monte Carlo study is performed on
purely Gaussian
data. We assume that the candidates have
an original $\satf$ value at the significance threshold $\satf = 4.77$ which translates into
$\rho^2_\text{cand} \approx 0.77$. Such a candidate represents
the lowest $\satf$ values considered in this search and will thus yield the highest false alarm. The
expected $2\F^{0.77}_\text{coh}$
value for
the coherent follow-up search for such candidates is $2\F^{0.77}_\text{coh} = 105.83$. Now,
$1\,000$
different realizations of pure Gaussian noise are created and searched with the coherent follow-up
search setup. In each of these the loudest candidate is identified and $\mathcal{R}$ is computed.
Figure~\ref{fig:falsealarm} shows the resulting distribution. None of the values exceeds
the threshold,
yielding a Gaussian false-alarm rate for this
veto of $\leq 0.1\%$. Of course the actual false-alarm rate would need to be measured on noise that
is
a more faithful representation of the actual
data rather than simple Gaussian noise. This is not trivial because one would need to characterize
the coherence properties of very weak disturbances in the data, which has never been
done.

Six of the 1\,138 tested candidates pass this veto in the \GCSearch{}, and all of them can be
ascribed to a hardware injection performed during S5 \cite{hardwareinjections,beritthesis}.

\section{Upper limits}
\label{lab:upperlimits}

No GW search to date has yet resulted in a direct detection. For CW
searches a null result is typically used to place upper limits on the
amplitude of the signals with parameters covered by the search. In
wide parameter-space searches these are typically standard frequentist upper limits (as they were
reported in \cite{Abbott:2003yq,Abbott2005a,Abbott2007a,Abbott2008a,Abbott2009c,Aasi:2012fw}) based
on the so-called ``loudest event.'' In most searches they are derived through intensive Monte Carlo
procedures. Only one search has used a numerical estimation procedure to obtain upper-limit values
\cite{Wette2008}. In
the \GCSearch{} a new, very cost-effective analytic procedure was used
which we describe here.

\subsection{The constant-$\eta$ sets}
\label{ch:constantchunks}

Generally, the search frequency band is divided into smaller {\it sets} and a separate upper limit
is derived for a population of signals with frequencies within each of these. The partitioning of
the parameter space can be done in different ways, with different advantages and disadvantages. Past
searches have often divided the parameter space into equally sized frequency bands. For example, the
Einstein@Home all-sky searches have divided the frequency band into $0.5$\,Hz-wide subbands, e.g.
\cite{Aasi:2012fw}. For a search like the \GCSearch{}, where the spin-down range grows with
frequency, such an approach would lead to significantly larger portions of parameter space for sets
at higher frequencies. Therefore, a slightly different approach is followed which divides the
parameter space into sets containing an approximately constant number of templates. The main
advantage of this approach is the approximately constant false-alarm rate over all sets. The total
number of templates $N$ is divided into $3\,000$ sets of $\sim\eta$ templates by subdividing the
total frequency range covered by the search into smaller subbands. This results in sets
small enough that the noise spectrum of the detectors is about constant over the frequency band of
each set. The number of templates in a set that spans a range $\Delta f = f_\Max - f_\Min$ in
frequency and $\Delta \dot f = \dot f_\Max - \dot f_\Min$ in spin-down is calculated with
\begin{align}\label{eq:sets}
\eta &= \eta_f \times \eta_{\dot f} = \frac{\Delta f}{\delta f}\times\frac{\Delta \dot f}{\delta
\dot f}
\end{align}
where the spacings are given in Eq.~\ref{eq:grid-spacings}. Hence, for a given minimum frequency
$f_\Min$, the maximum frequency $f_\Max$ associated with a set is
\begin{align}
f_\text{max} &= \frac{f_\text{min}}{2} + \sqrt{\left(\frac{-f_\text{min}}{2}\right)^2 + \eta\ (200\
\ys)\ \delta f\ \delta\dot f}.
\end{align}
The total number of templates searched in the \GCSearch{}, $N \simeq 4.4 \times 10^{12}$,
is in this way divided into the $3\,000$ sets, each containing $\eta \simeq 1.5\times 10^{9}$
templates. 

Because of known detector artifacts in the data (see Sec.~\ref{lab:knownlines}), not each of these
sets is assigned an upper-limit value. Some sets entirely comprise frequency bands excluded from the
postprocessing by the known-lines cleaning procedure.  We cannot make a statement about the
existence of a GW
signal in such sets and, hence, no upper-limit value is assigned to those sets. Other
sets are only partially affected by the known artifacts and an upper-limit value can still be
assigned on the {\it considered} part of the parameter space. However, in order to keep the
parameter-space volume associated with each set about constant, upper-limit values are assigned only
on sets with a relatively low fraction of the {\it excluded} parameter space, as will be explained
below.
\par
\begin{figure}
	\includegraphics[trim = 0mm 30mm 8mm 0mm, clip, width=0.5\textwidth]{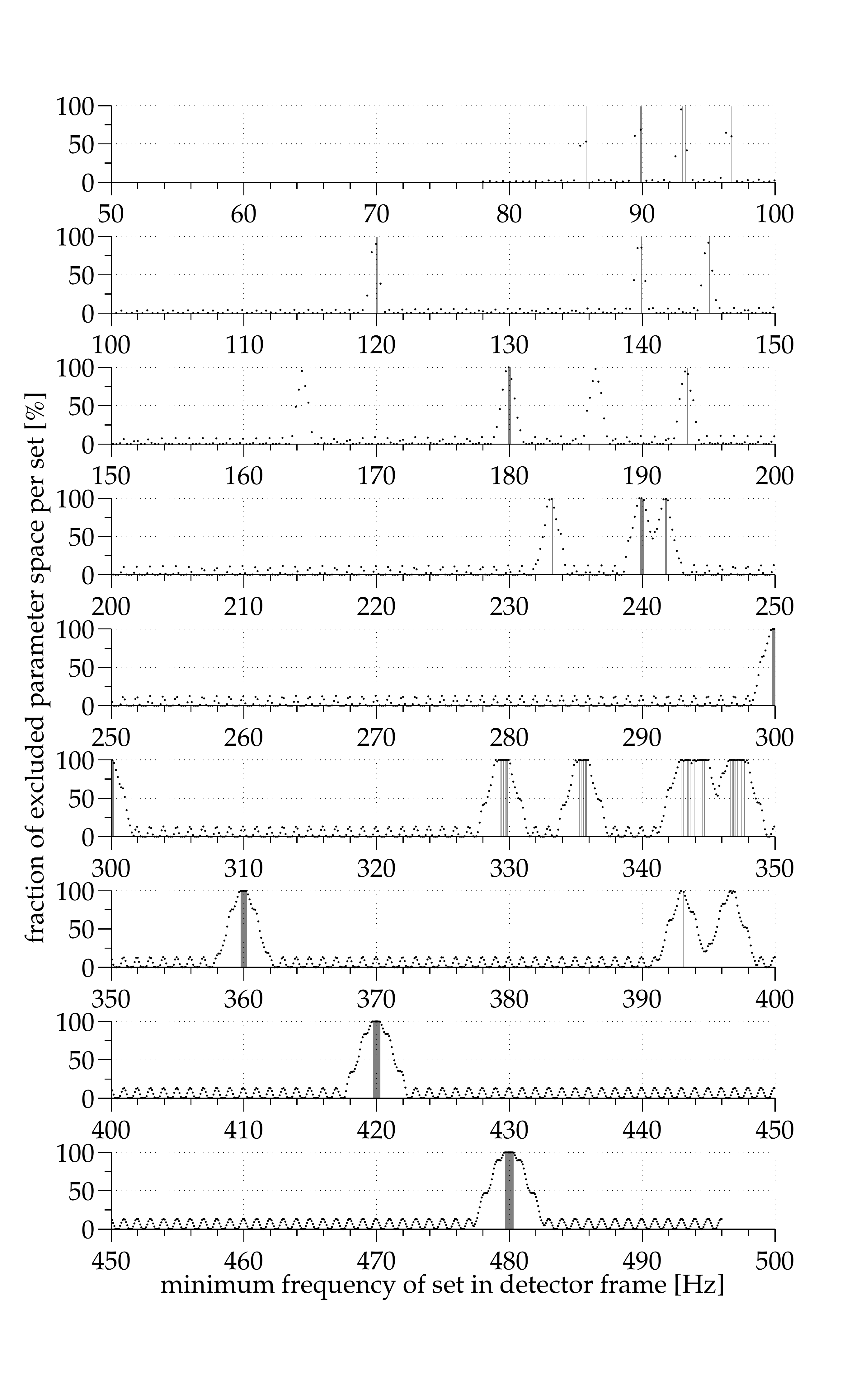}
	\caption{Here we show the percentage of excluded parameter space per upper-limit set as a
function of frequency.
	The gray vertical lines
denote the frequencies that host known detector artifacts and which are cleaned strictly. The
1\,Hz lines are not shown. The periodical variations of the excluded parameter space are due to the
$1\,\Hz$ harmonics. We remind the reader that the searched frequency band starts at $78$\,Hz and
ends at
$496$\,Hz.
}
	\label{fig:lostspace}
\end{figure}

Figure~\ref{fig:lostspace} gives an overview of the excluded parameter space per set. Each black
data
point denotes the amount of excluded parameter space for one set. The additional red lines mark the
known spectral artifacts of the detector that are vetoed strictly (the 1\,Hz lines are not
shown). The $60$\,Hz power lines are clearly visible, as well as,
for example, the calibration line at $393.1$\,Hz (compare Tables VI and VII in \cite{Aasi:2012fw}).
The
effect of the presence of the $1\,\Hz$ harmonics is visible throughout the whole frequency range.

\begin{figure}
	\includegraphics[width=0.5\textwidth]{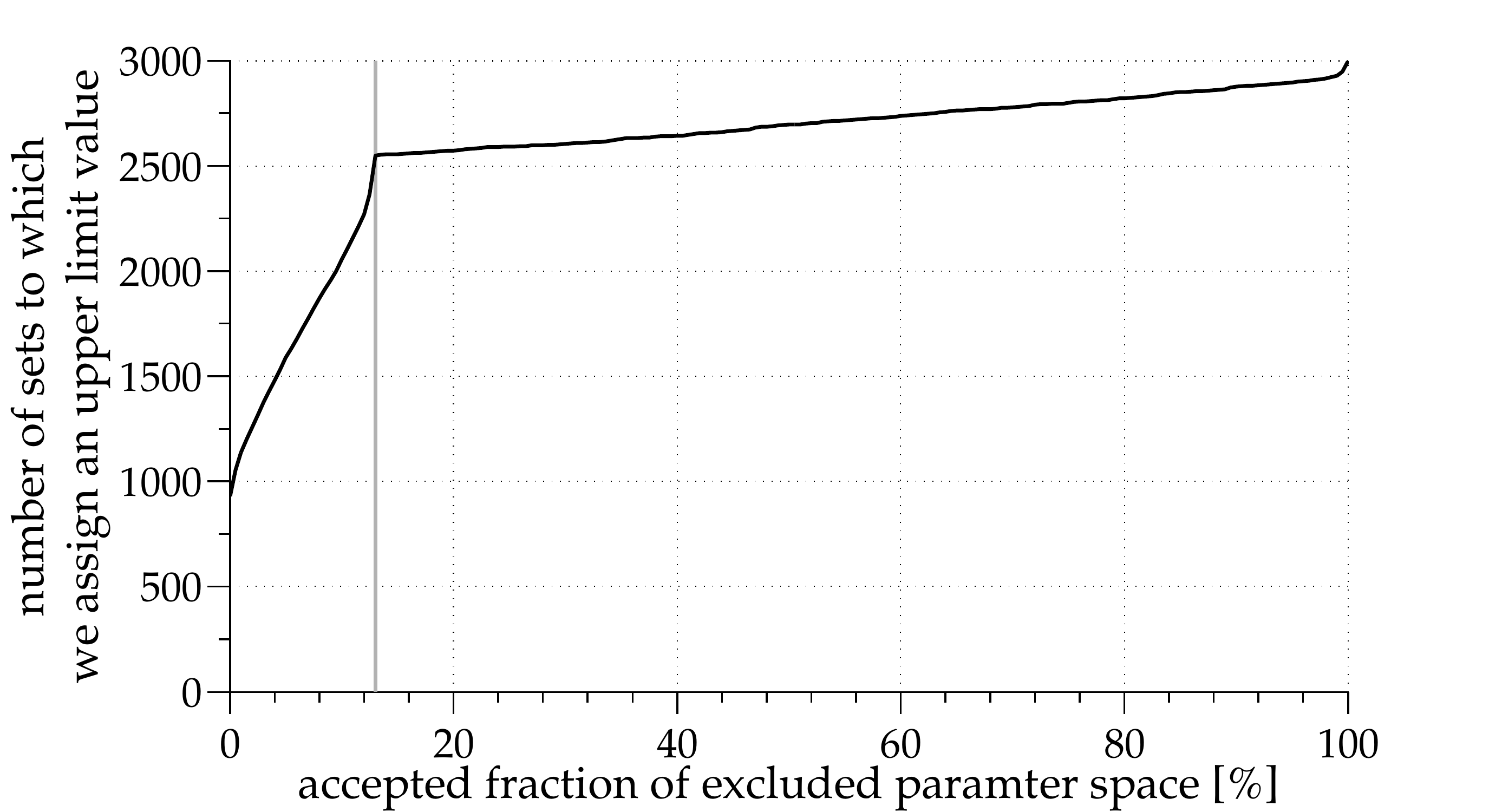}
	\caption{The cumulative distribution of the excluded parameter space per set (black line). The
distribution shows a steep increase towards the $13\%$ threshold (gray line). After reaching that
threshold, the distribution has a shallow knee up to the $100\%$ loss. As explained in the text, the
shape of this
distribution simplifies the choice of the threshold.}
	\label{fig:validthreshold}
\end{figure}

An upper-limit value is assigned to all sets for which the excluded parameter space is not more than
$13\%$ of the total. Figure~\ref{fig:validthreshold} shows the cumulative distribution of the
excluded parameter space of the $3\,000$ sets.
The shape of the
distribution clearly suggests picking the threshold at $13\%$, which keeps the maximum
number of segments while minimizing the excluded parameter space per segment. As a
result, an upper-limit value is placed on $2\,549$ sets.

\subsection{The CW loudest-candidate upper limit}
\label{ch:standardprocedure}

The standard frequentist upper limit is the intrinsic gravitational-wave amplitude such that a high
fraction (90\% or 95\%) of the population of signals being searched, at that amplitude, would have
yielded a value of the detection statistic higher than the highest one that was produced by the
search (including any postprocessing). The direct way to determine the upper limit value consists
of injecting a certain number
[$\mathcal{O}(100)$] of signals into the original data set used for the search. The injected signals
have parameters randomly distributed within the searched parameter space. All signals are injected
at fixed amplitude $h_0$. A small region around the injections in frequency and spin-down (and
for the all-sky searches also in sky) is analyzed and the loudest candidate is identified. This
candidate then undergoes the complete postprocessing pipeline. If it survives all vetoes, and if its
$\satf$ value exceeds the $\satf$ of the most significant surviving candidate from the search, then
such an injection is counted as {\it recovered}. The fraction of all recovered injections to the
total
number of injections gives the {\it confidence} value $C(h_0)$. The $h_0$ value associated
with, say, $90\%$ confidence is $h_0^{90\%}$. If a signal of that strength had been present in the
data, $90\%$ of the possible signal realizations would have resulted in a more significant candidate
than the loudest that was measured. Thus, the presence of a signal of strength $h_0^{90\%}$ or
louder in the data set is excluded and $h_0^{90\%}$ is the $90\%$ confidence upper-limit value on
the GW amplitude.

A different $h_0^{90\%}$ can be assigned to different portions of the searched parameter space. It
will depend on the loudest candidate from that portion of parameter space, on the noise in that
spectral region and on the extent of the parameter space. In general, in order to derive the upper
limit
for each portion of parameter space, several injection sets at
different $h_0$ values have to be carried out in order to bracket the desired confidence level.
Ultimately, an interpolation can be utilized to estimate $h_0^{C}$. This standard approach is
extremely time consuming because it requires the production and search of several hundred fake data
sets for each portion of parameter space on which an upper limit is placed.

\subsection{Semianalytic upper limits}

The basic idea is that it is not necessary to sample the different $h_0$ values with different
realizations of the noise and polarization parameters. Instead, the same noise and signal
realizations
can be reused for different $h_0$ values to sample at virtually no cost the $C(h_0)$ function and
find the upper-limit value at the desired confidence level. \par
The relation between the measured $\satf$ value and the injection strength $h_0$ can to good
approximation be described by the following relation:
\be
	\satf \simeq \langle \mathcal{N}\rangle +  \langle \mathcal{G}\rangle  h_{0}^2,
	\label{eq:semianalytic}
\ee
where $\langle \mathcal{N} \rangle$ represents the average contribution of the noise to the
detection
statistic value for a given putative signal and $\langle \mathcal{G}\rangle$ is proportional to the
average contribution of the signal and depends on the signal parameters, on the time\-stamps of the
data and on the noise-floor level. We fix the parameters that define a signal and we
obtain $\langle\mathcal{N}\rangle$ and $\langle \mathcal{G}\rangle$ from the $\satf$ values measured
by injecting two signals with a
different $h_0$ value, keeping {\it all} other parameters fixed. With this information it is
possible to estimate $\satf$ for any value of $h_0$ for that particular combination of signal
parameters and data set. With two sets of $N_t$ injections and searches we
produce $N_t$ $\{\langle\mathcal{N}\rangle,\langle\mathcal{G}\rangle\}$ couples, corresponding to
$N_t$ signals. We use these $\{\langle \mathcal{N}\rangle,\langle\mathcal{G}\rangle \}$
couples to predict $N_t$ values of $\satf$ for any $h_0$ with Eq.~\ref{eq:semianalytic}. From these
the confidence is immediately estimated by simply counting how many exceed the loudest measured one,
without further injection-and-search studies.

This semianalytic upper-limit procedure requires only two cycles through the
in\-jec\-tion-and-search procedure per signal, corresponding to two different signal strengths,
namely  $5 \times 10^{-25}$ and $7\times
10^{-25}$ for the \GCSearch{}. With $2\,549$ upper-limit sets, each requiring $N_t=100$ signals,
this results
in $509\,800$ injections. In some especially noisy sets, more than $100$ injections are
performed amounting to a total of $796\,400$ injections. Each job needs about 20~min to create
and search the data. Assuming $\sim 1\,000$ jobs running in parallel on the ATLAS compute cluster,
the whole procedure takes a few days up to a week. This is significantly less than the time needed
by the standard approach (days as opposed to weeks).

The main advantage of our method is that the injections can be made with arbitrary signal
amplitudes, while the
standard approach requires an educated guess of $h_0$ to start with. The standard
in\-jec\-tion-and-search procedure is then repeated for
different $h_0$ values until the 90\% confidence upper limit-value is found. In principle, once the
first
$h_0^{90\%}$ value is found, one can rescale it by the noise level to estimate the $h_0$ upper limit
of neighboring
sets. But these values are oftentimes not correct, because of noise fluctuations in the data within
one set.
Experience has shown that at least five to ten different $h_0$
values need to be explored for each band, resulting in a computational effort several times larger
than that of our semianalytic procedure.

\begin{figure}
	\includegraphics[width=0.5\textwidth]{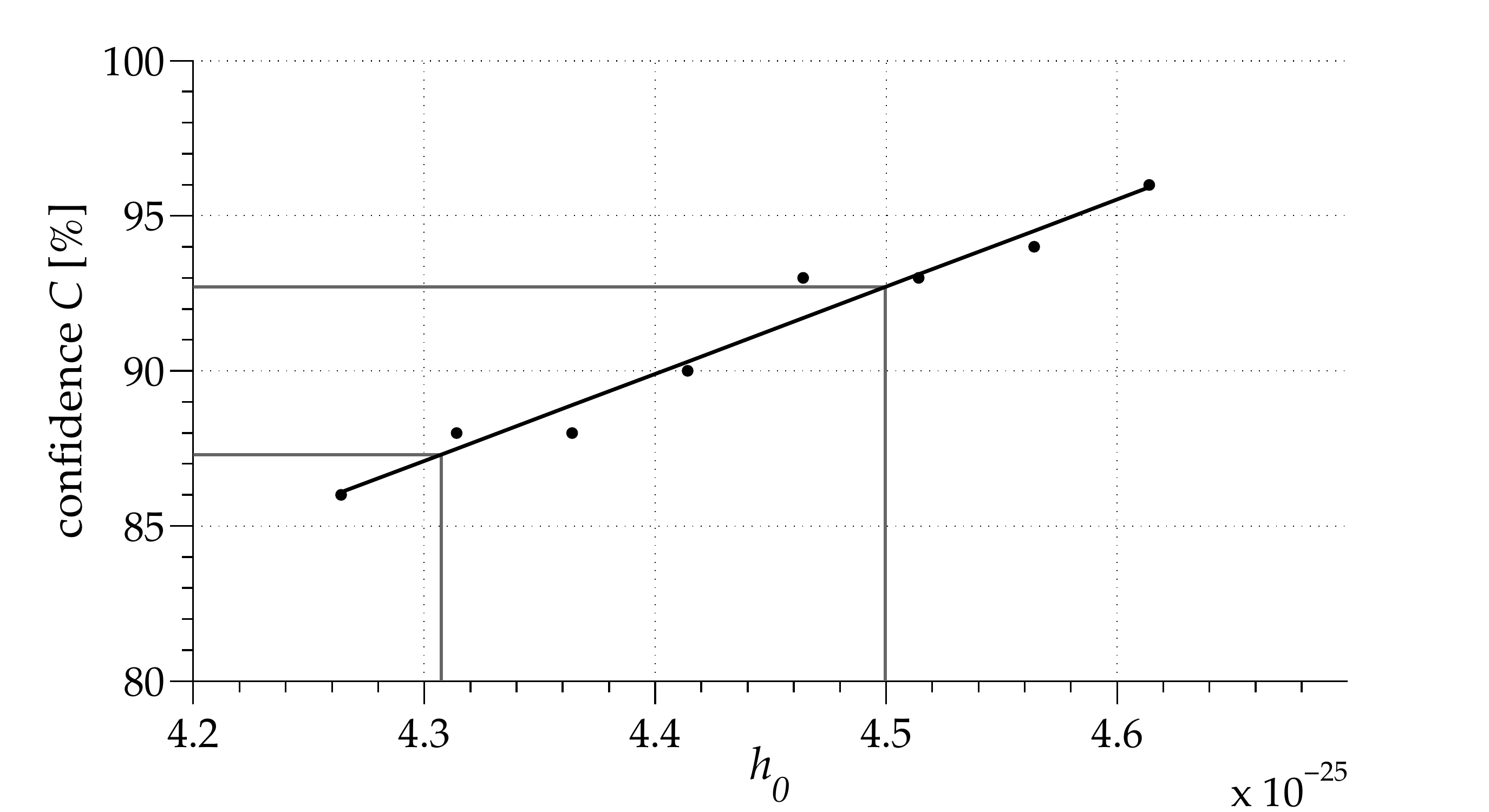}
	\caption{The plot shows $C(h_0)$ for different $h_0$ values for one
example set. By fitting a straight line to the data points in a small enough region around the
$90\%$ confidence value the $3\%$ error on the confidence can be translated into an uncertainty of
$\lesssim 5\%$ on $h_0^{90\%}$.}
	\label{fig:ulexample}
\end{figure}

To estimate the uncertainty on the upper-limit values we use a linear approximation to the curve
$C(h_0)$ in the neighborhood of $h_0^{90\%}$. Figure~\ref{fig:ulexample} shows $C(h_0)$ in that
region for a set at about $150$\,Hz. The $1\sigma$ uncertainty in $C(h_0)$ based on the binomial
distribution for $100$ trials with single-trial probability of success $90\%$, is $3\%$.
As illustrated in Fig.~\ref{fig:ulexample}, this corresponds to an uncertainty of $\lesssim 5\%$ on
$h_0^{90\%}$.

The choice of $N_t=100$ allows significant savings with respect to the standard upper-limit
procedure and also yields statistical uncertainties which are acceptable ($5\%$ is less than the
typical uncertainty in the calibration amplitude which is typically $\sim 10\%$). However, one may
still wonder whether 100
signals are a representative sample of the signal population in the parameter-space volume that the
upper limit
refers to. To prove this we have compared the results of the semianalytic upper-limit procedure
for different numbers of trials from $N_t=10$ to $N_t=1\,000$. The results of this study confirm the
5\% error on $h_0^{90\%}$. A larger number of trials leads to a reduced uncertainty, like $\sim 2\%$
for $N_t=200$ and $< 1\%$ for $N_t=1\,000$.

Finally, we validate the results of the semianalytic procedure using the standard
injection-and-search procedure on
a sample of sets. For this we inject 100 signals corresponding to the signal population described in
Table~\ref{tab:falsedismissalparams} into the original search data set. The data are analyzed in a
small template grid around the injection (a subset of the original grid) and the loudest candidate
is identified. The confidence value is obtained by counting in what fraction of the 100 trials this
candidate has a higher $\satf$ value than the loudest remaining candidate in that set.
Table~\ref{tab:confverification} shows
the results for ten different, randomly chosen sets. In all but two cases the measured confidence
lies within $1\sigma$ statistical uncertainty value based on 100 trials. In the other two sets the
confidence is larger than $90\%$ plus one standard deviation.

\begin{table}
    \centering
  \begin{tabular*}{0.48\textwidth}{rrccr}
	\hline
    Set ID & $~f_\Min^\text{set}$[Hz]~& $h_0^{90\%}[\times 10^{-25}]~$ & Threshold &
$~~C(h_0^{90\%})$[\%]  \\
	\hline
	$57$ & $103.12$ & $5.85$ & $4.7523$ & $93$ \\
	$254$ & $162.42$ & $3.67$ & $4.7098$ & $92$ \\
	$363$ & $187.34$ & $3.80$ & $4.7098$ & $88$ \\
	$631$ & $237.77$ & $4.50$ & $4.7394$ & $92$ \\
	$1025$ & $296.74$ & $5.52$ & $4.7497$ & $90$ \\
	$1586$ & $364.61$ & $7.13$ & $4.7446$ & $90$ \\
	$1672$ & $373.93$ & $6.75$ & $4.6984$ & $90$ \\
	$2302$ & $436.16$ & $7.77$ & $4.7218$ & $93$ \\
	$2695$ & $470.83$ & $8.49$ & $4.7303$ & $94$ \\
	$2972$ & $493.81$ & $11.8$ & $4.7170$ & $100$ \\
    \hline
  \end{tabular*}
	\caption{Validation of the $h_0^{90\%}$ values obtained with the semianalytic procedure for ten
sample sets. For each set $100$ GW
signals have been injected at the $h_0^{90\%}$ level into the search data set. The confidence is the
fraction of these injections that was recovered with a higher $\satf$ value than that of the loudest
surviving candidate of that set.}
	\label{tab:confverification}
\end{table}

We note that in the upper-limit procedure we do not subject the injections to any of the
postprocessing vetoes (including the follow-up) that we apply to the search. This is because of the
very low false-dismissal rate of these vetoes.
In $98.1\%$ of all cases in which a signal candidate
is not recovered, it is because the candidate fails the comparison with the loudest surviving
candidate from the search in that set. Only in $0.2\%$ of the cases is it the $\F$-statistic
consistency veto that discards the candidate and in $1.7\%$ the candidate is lost at the
permanence veto step.  This systematic loss of detection efficiency was neglected in
the \GCSearch{} and it would have resulted in a 3\% increase in the upper-limit values, as can be
seen from Fig.~\ref{fig:ulexample}.

%
%
%
\section{Second-order spin-down}
\label{lab:secondspindown}

All methods described in the previous sections have been assessed on a population of signals without
second-order spin-down. However, if the observation time is of the order of years, the second-order
spin-down has an impact in the incoherent combination of the single-segment results. The minimum
second-order spin-down signal that is necessary to move the signal by a frequency bin $\delta f$
within an observation time $T_\text{obs}$ is
\be\label{eq:safessd}
\ddot f_\Min = \frac{\delta f}{T_\text{obs}^2}.
\ee
In case of the \GCSearch{} this value is $\sim 6\times 10^{-21}\,\Hz/\s^2$.

To quantify the false-dismissal for signals with second-order spin-down we perform an
injection-and-search Monte Carlo study where we inject signals with a second-order spin-down and
search for them and perform postprocessing as described above, namely without taking account of a
second-order spin-down.
A set of $500$ signal injections is created with the parameters given in
Table~\ref{tab:falsedismissalparams} and with
an additional second-order spin-down value in the range $0 \leq \ddot f \leq n \dot f^2 /f$
\cite{Palomba:1999su} with a
braking index of $n=5$ \cite{Wette2010}. Since the second-order spin-down values are drawn from a
uniform random distribution, most of them fall in the higher range ($\ddot f \sim
10^{-17}\,\Hz/\s^2$) and this
study is representative of the worst case (high second-order spin-down) scenario. The data are
analyzed using the original template setup of
the corresponding job of the search and the highest $\satf$ value within a region as large as the
cluster size around the injection is recovered. The 36 signals would not have been recovered by the
original search because they would not have been included in the top 100\,000 that
were recorded. Of the remaining $464$ candidates from the injections 458 survive the $\F$-statistic
consistency veto, which implies a false-dismissal rate of $\sim 1.3\%$. The next step is the
selection of the most significant subset with a significance threshold at $\satf \ge 4.77$. A total
of 322
($\sim
70.3\%$) of the signals are above this threshold. Then, the permanence veto is applied and
nine signals are lost, corresponding to a false-dismissal rate of $\sim 2.8\%$. Overall, the
postprocessing steps up to the coherent follow-up amount to a loss of signals of about 37\%, mostly
due to the application of the significance threshold. However, as mentioned above, this loss is the
result of a worst case study. We find empirically that at 
$\ddot f \leq 5\times 10^{-20}\,\Hz/\s^2$ 
the additional second-order spin-down component
does not
change the false-dismissal rates and upper-limit results reported above (for the targeted signal
population).

Obtaining the false-dismissal rates for a variety of different second-order spin-down values
for the $\R$ veto in the coherent follow-up search is
computationally very demanding. Therefore, we restrict our study to a few second-order spin-down
values.
The false-dismissal remains very low for second-order spin-down signals with values lower or
equal to $5\times 10^{-20}\,\Hz/\s^2$. In a study over $1\,000$ injections no candidate was lost,
hence, the false-dismissal rate is $\leq 0.1\%$. In contrast, the false-dismissal rate for a signal
population with second-order spin-down values of $4\times 10^{-17}\,\Hz/\s^2 \leq \ddot f \leq
5\times 10^{-17}\,\Hz/\s^2$ is about $ 82\%$.
Based on this, we expect our search to be significantly less sensitive for signals with a second
order spin down larger than $5\times 10^{-20}\,\Hz/\s^2$. We find that the 90\%-confidence upper
limit
values for a population with second-order spin-down values
of order $\sim 10^{-17}\,\Hz/\s^2$ are about a
factor of $2$ higher than the upper-limit values presented in the \GCSearch{}.

%
%
\section{Discussion}
\label{lab:discussion}

The final sensitivity of a search depends on the applied search methods, on the searched parameter
space and on the data set used.
In order to be able to quantify and compare the sensitivity of different
\emph{searches},
independently of the quality and sensitivity of data used, we introduce the \emph{sensitivity depth}
$\cD^\text{C}$ of a search as
\be\label{eq:searchdepth}
\cD^{C} (f) \equiv \frac{\sqrt{\Sh (f)}}{h^{C}_0(f)},
\ee
where ${h^\text{C}_0(f)}$ is the upper limit (or amplitude sensitivity estimate) of confidence $C$
and $\Sh(f)$ is the noise power spectral density at frequency $f$.
The average sensitivity depth of the \GCSearch{} is found as $\cD^{90\%} \simeq 75\,\Hz^{-1/2}$ in
the band
between 100 and 420\,Hz and drops somewhat at higher and lower frequencies.
For comparison, the sensitivity depth of the last Einstein@Home search \cite{Aasi:2012fw}
was $28\,\Hz^{-1/2} \leq\cD^{90\%} \leq 31\,\Hz^{-1/2}$ and
that of the coherent search \cite{Wette2010} for signals from the compact source in Cassiopeia~A
was $\cD^{95\%}
\simeq 37\,\Hz^{-1/2}$.

The high sensitivity depth of the \GCSearch{} compared to these searches is related to the amount of
data used, the fact that it is a directed search, and the particular search setup and
postprocessing.
In this paper we present the search setup and postprocessing techniques developed to achieve such
high sensitivity.
In the following we briefly summarize and discuss the main results.



We present a data-selection criterion that has not been used in previous GW searches.
We compare this selection criterion with three
other methods from the literature
and show that, for a search like the \GCSearch{}, our selection criterion results in the
most sensitive data set, improving the overall sensitivity by $20\%-30\%$ with respect to the next
best data-selection criterion (used in the Scorpius X-1 search \cite{Abbott2007a}).
An analytical estimator exists for our criterion which avoids the numerical simulation of signals
with different polarization parameters, namely,
\be Q_{\cD} = \frac{2}{5}h_0^2 \,Q_{\cD'} + 4\ee
with
\begin{equation}\label{ourfigureofmerit}
  Q_{\cD'} = \sum_{k=1}^{N_\seg^\SFT}\frac{\mathcal{P}^k}{\Sh^k}.
\end{equation}
The techniques to remove nonsignal candidates presented in this paper have allowed us to
probe the \GCSearch{} candidates associated with values of the detection statistic that are
marginal over the parameter space of the original search.
These techniques drastically reduce the number of candidates, while at
the same time they are safe against falsely dismissing real signals. We have presented the
various techniques in the order that they were applied. This order was principally dictated by
trying to optimize the depth of the search (i.e. maximize the number of candidates that could be
meaningfully
followed up) at a constrained computational cost.
An overview of the effectiveness in the example of the \GCSearch{} is given in
Table~\ref{tab:postprocoverview} and Fig.~\ref{fig:summaryplot} and below we review the main steps.
In the \GCSearch{}, six candidates survive the whole postprocessing. They can
all be ascribed to a hardware injection that was performed during S5.

\begin{itemize}
\item[(i)] At first, we remove candidates that could stem from {\it known detector artifacts}. This
technique has been used in
past searches. Here we introduce a variant, namely, a relaxed
cleaning procedure, in order to deal with the frequent occurrence of the $1\,\Hz$ line harmonics.
Recently, a generalization of
the $\F$ statistic has been developed, which is more
robust against noncoincident lines in multiple detectors \cite{Keitel:2013wga} than the $\F$
statistic. It will be interesting to compare the performance of these different approaches.
\item[(ii)] Candidates that can be ascribed to the same tentative signal are combined and only a
single
representative candidate is kept. With this {\it clustering} procedure the computational effort of
the
subsequent steps is reduced (by reducing the number of candidates to investigate).
A further development of this clustering scheme consists of computing a representative
point in parameter space by weighting the statistical significance of the candidates within a
cluster box, instead of picking the loudest.
This approach is currently being developed for the clustering of candidates from all-sky
Einstein@Home searches.


\begin{table}[t]
    \centering
  \begin{tabular*}{0.48\textwidth}{lrlc}
	\hline
	Applied veto & Candidates left & \phantom{0} &  [\%] \\
	\hline
	Analyzed number of templates \ \ & $4.4\times 10^{12}$ & & 100\\
	Reported candidates & $1\times 10^9$ & &$0.024$\\
	Known-lines cleaning & $8.9\times 10^8$ & & $83$\\
	Clustering & $2.9\times 10^8$ & & $33$ \\
	$\F$-statistic consistency veto & $2.6\times 10^8$& &$88$\\
	Significance threshold & $27\,607$ & & $0.01$\\
	Permanence veto & $1\,138$ & & $4$\\
	Coherent follow-up search & $6$ & & $0.5$ \\
    \hline
  \end{tabular*}
  \caption{The number of candidates at each stage of the search. The first column
indicates the name of the stage, the second column shows
the number of candidates surviving that stage; the third column gives the fraction of
candidates that survive a particular stage with respect to the number of candidates that
survived the previous stage.}
  \label{tab:postprocoverview}
\end{table}

\begin{figure}[t]
	\includegraphics[width=0.5\textwidth]{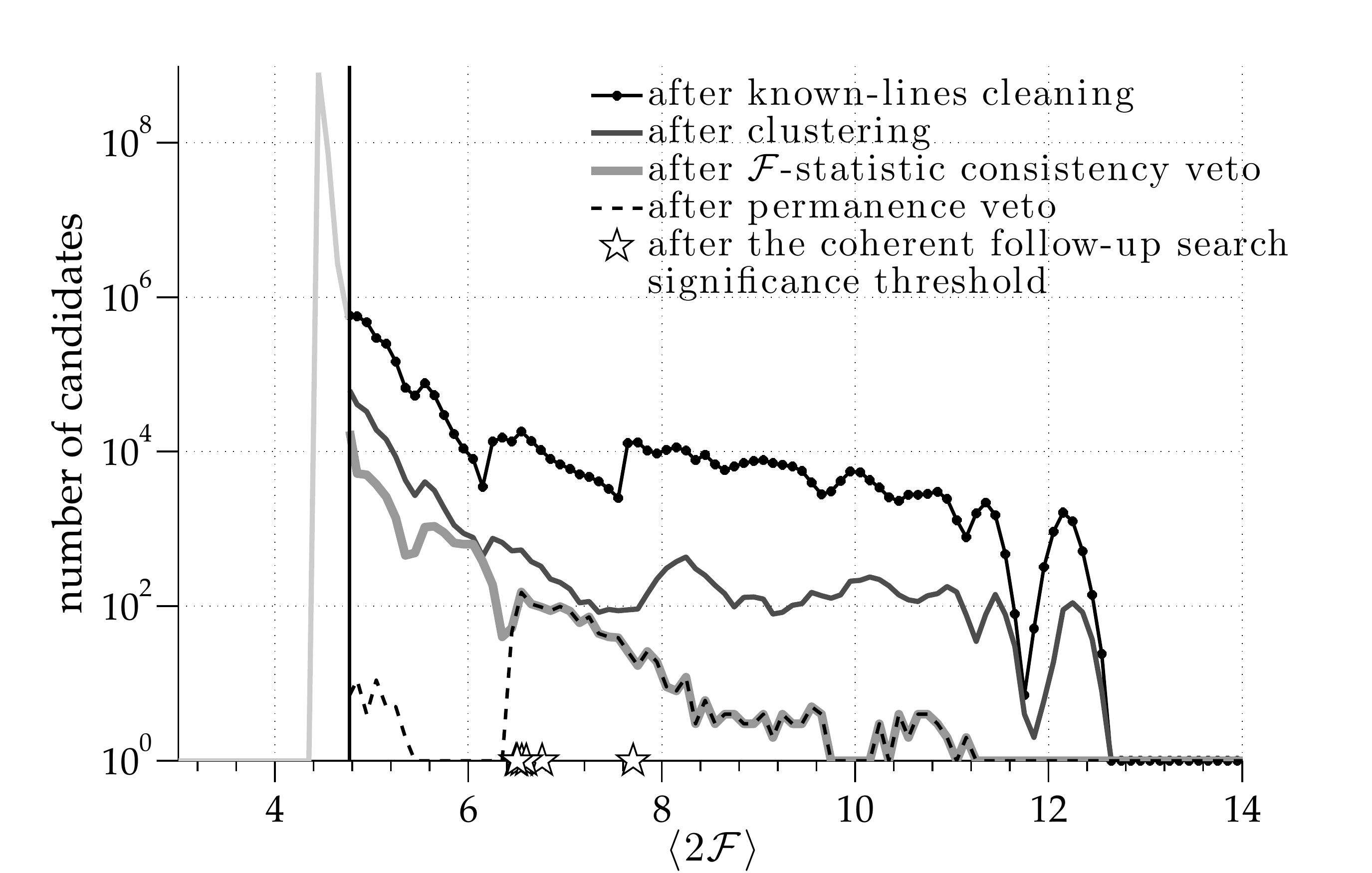}
	\caption{The plot shows the distributions of the $\satf$ values of the surviving candidates
after each of the postprocessing steps. The vertical solid line denotes the significance threshold
at $4.77$. The dotted line denotes the candidates that survive the known-lines cleaning. The
clustering
(solid, dark gray) lowers the histogram evenly over the different $\satf$ values, as one would
expect it. The
$\F$-statistic consistency veto (bold, gray) removes high $\satf$ outliers and lowers the right
tail.
The application of the permanence veto removes many of the low-significance candidates. All
candidates with $\satf$ values above $\sim 6$ survive this step. However, these can all be ascribed
to a hardware injection that was performed during S5. The remaining group of $59$ ``real'' CW
candidates are then ruled out by the coherent follow-up
search and the $\mathcal{R}$ veto. Six of the candidates that can be ascribed to the hardware
injection would pass
this veto.}
	\label{fig:summaryplot}
\end{figure}

%

\item[(iii)] The {\it $\F$-statistic consistency veto} removes candidates whose multidetector
$\F$-statistic value is lower than any of the single-detector values.
This procedure has already been used on former searches (see, for example
\cite{Aasi:2012fw}) and also generalized in  \cite{Keitel:2013wga}.

\item[(iv)] {\it Permanence veto:} Candidates that display an accumulation of significance in a very
short
time rather than a constant rate of accumulation over the entire observing time are removed from the
list of possible CW signals. This veto has not been used in former searches and shows high
effectiveness and a low false-dismissal rate for strictly continuous GW signals.
We note however that there may be transient CW signals in the data, lasting days to weeks, that this
step might dismiss.
A follow-up on the dismissed candidates, specifically targeting transient CW signals, is an
interesting future project.

\item[(v)] A {\it coherent follow-up} search is performed with fine template grids (and low
mismatch)
around each candidate and over a time span of
$90$ days. This is a different and simpler approach
than the one taken in \cite{Aasi:2012fw} and detailed in \cite{shaltev:2014} which
uses two stages and a nondeterministic coherent follow-up strategy. The simple criterion chosen to
discard
candidates based on the outcome of the coherent follow-up is illustrated and its effectiveness and
safety are demonstrated.
\end{itemize}

The analytical upper-limit procedure described in this paper significantly reduces the computational
effort with respect to the standard frequentist upper-limit procedure. For most upper-limit
subsets of the parameter space only two inject-and-search cycles with $N_t=100$ are necessary. Only
if the data used
in a set is particularly noisy are higher values of $N_t$ necessary in order to adequately
characterize the possible outcomes of the search in that set. The upper-limit derivation of the
\GCSearch{} took only about a week on $\sim 1\,000$ nodes of the ATLAS compute cluster, which is
significantly less than the standard procedure (see, for example, reference
\cite{Abbott:2003yq}).

%
%
\begin{acknowledgments}
We are grateful to the members of the continuous waves working group at the Max Planck Institute for
Gravitational Physics, in particular to Badri Krishnan and Oliver Bock for their continued support
of this project, as well as David Keitel.
We would also like to thank the members of the continuous waves working group
of the LIGO-Virgo Collaboration for helpful comments and discussions, particularly Pia Astone and
Ben Owen.
This paper was reviewed by Siong Heng on behalf the LIGO Scientific Collaboration: his insightful
comments and thorough
reading have improved it and we are very grateful for the time that he spent on it. We gratefully
acknowledge the support of the International Max Planck Research School on Gravitational-Wave
Astronomy of the Max Planck Society and of the Collaborative Research Centre SFB/TR7 funded by the
German Research Foundation. This document was assigned LIGO Document No. P1300125 and AEI Document
No. AEI-2014-017.
\end{acknowledgments}

\bibliographystyle{unsrt}
\bibliography{Bibliography}

\begin{thebibliography}{10}

\bibitem{myobspaper}
J.~Aasi et~al.
\newblock {Directed search for continuous gravitational waves from the Galactic
  center}.
\newblock {\em Phys.\ Rev.\ D}, 88:102002, 2013.

\bibitem{lalsuite}
{LIGO Scientific Collaboration}.
\newblock Lal/lalapps software suite.
\newblock \url{http://www.lsc-group.phys.uwm.edu/daswg/projects/lalsuite.html}.
\newblock Accessed: 06/03/2013.

\bibitem{jks}
Piotr Jaranowski, Andrzej Kr\'olak, and Bernard~F. Schutz.
\newblock Data analysis of gravitational-wave signals from spinning neutron
  stars: The signal and its detection.
\newblock {\em Phys.\ Rev.\ D}, 58(6):063001, 1998.

\bibitem{Cutler:2005hc}
Curt Cutler and Bernard~F. Schutz.
\newblock {The Generalized $\mathcal{F}$-statistic: Multiple detectors and
  multiple GW pulsars}.
\newblock {\em Phys.\ Rev.\ D}, 72:063006, 2005.

\bibitem{Prix:2006wm}
R.~Prix.
\newblock {Search for continuous gravitational waves: Metric of the
  multidetector $\mathcal{F}$-statistic}.
\newblock {\em Phys.\ Rev.\ D}, 75(2):023004, 2007.

\bibitem{Brady1998b}
Patrick~R. Brady and Teviet Creighton.
\newblock {Searching for periodic sources with LIGO. II: Hierarchical
  searches}.
\newblock {\em Phys.\ Rev.\ D}, 61:082001, 2000.

\bibitem{PhysRevD.70.082001}
Badri Krishnan, Alicia~M. Sintes, Maria~Alessandra Papa, Bernard~F. Schutz,
  Sergio Frasca, and Cristiano Palomba.
\newblock Hough transform search for continuous gravitational waves.
\newblock {\em Phys.\ Rev.\ D}, 70:082001, 2004.

\bibitem{Pletsch2009}
Holger~J. Pletsch and Bruce Allen.
\newblock Exploiting large-scale correlations to detect continuous
  gravitational waves.
\newblock {\em Phys.\ Rev.\ Lett.}, 103(18):181102, 2009.

\bibitem{Wette2008}
K.~{Wette} et~al.
\newblock {Searching for gravitational waves from Cassiopeia A with LIGO}.
\newblock {\em Class. Quant. Grav.}, 25(23):235011, 2008.

\bibitem{PrixShaltev:2012}
Reinhard Prix and Miroslav Shaltev.
\newblock {Search for Continuous Gravitational Waves: Optimal StackSlide method
  at fixed computing cost}.
\newblock {\em Phys.\ Rev.\ D}, 85:084010, 2012.

\bibitem{Wette:2013wza}
Karl Wette and Reinhard Prix.
\newblock {Flat parameter-space metric for all-sky searches for
  gravitational-wave pulsars}.
\newblock {\em Phys.\ Rev.\ D}, 88:123005, 2013.

\bibitem{Abbott2009e}
B.P. Abbott et~al.
\newblock {LIGO: The Laser Interferometer Gravitational-wave Observatory}.
\newblock {\em Rept.\ Prog.\ Phys.}, 72:076901, 2009.

\bibitem{Abbott2009d}
B.~Abbott et~al.
\newblock {Einstein@Home search for periodic gravitational waves in early S5
  LIGO data}.
\newblock {\em Phys.\ Rev.\ D}, 80(4):042003, 2009.

\bibitem{Abadie:2010px}
J.~Abadie et~al.
\newblock {Calibration of the LIGO Gravitational Wave Detectors in the Fifth
  Science Run}.
\newblock {\em Nucl.~Instrum.~Meth.}, A624:223--240, 2010.

\bibitem{Abbott2009a}
B.~Abbott et~al.
\newblock {Einstein@Home search for periodic gravitational waves in LIGO S4
  data}.
\newblock {\em Phys.\ Rev.\ D}, 79(2):022001, 2009.

\bibitem{Wette2010}
J.~Abadie et~al.
\newblock {First search for gravitational waves from the youngest known neutron
  star}.
\newblock {\em Astrophys. J.}, 722:1504--1513, 2010.

\bibitem{Abbott2007a}
B.~Abbott et~al.
\newblock {Searches for periodic gravitational waves from unknown isolated
  sources and Scorpius X-1: Results from the second LIGO science run}.
\newblock {\em Phys.\ Rev.\ D}, 76(8):082001, 2007.

\bibitem{Aasi:2012fw}
J.~Aasi et~al.
\newblock {Einstein@Home all-sky search for periodic gravitational waves in
  LIGO S5 data}.
\newblock {\em Phys.\ Rev.\ D}, 87:042001, 2013.

\bibitem{prixetal2011:_transientCW}
Reinhard Prix, Stefanos Giampanis, and Chris Messenger.
\newblock Search method for long-duration gravitational-wave transients from
  neutron stars.
\newblock {\em Phys.\ Rev.\ D}, 84:023007, 2011.

\bibitem{2011PhRvD..83h3004T}
E.~{Thrane}, S.~{Kandhasamy}, C.~D. {Ott}, W.~G. {Anderson}, N.~L.
  {Christensen}, M.~W. {Coughlin}, S.~{Dorsher}, S.~{Giampanis}, V.~{Mandic},
  A.~{Mytidis}, T.~{Prestegard}, P.~{Raffai}, and B.~{Whiting}.
\newblock {Long gravitational-wave transients and associated detection
  strategies for a network of terrestrial interferometers}.
\newblock {\em \prd}, 83(8):083004, 2011.

\bibitem{hardwareinjections}
V.~Mandic and P.~Shawhan.
\newblock S5 hardware injections.
\newblock {\em LIGO Document LIGO-G060435-00-D}, 2006.

\bibitem{beritthesis}
Berit Behnke.
\newblock {\em {A directed search for continuous gravitational waves from
  unknown isolated neutron stars at the galactic center}}.
\newblock PhD thesis, Leibniz Universit\"at Hannover,
  http://edok01.tib.uni-hannover.de/edoks/e01dh13/752172840l.pdf, 2013.

\bibitem{Abbott:2003yq}
B.~Abbott et~al.
\newblock {Setting upper limits on the strength of periodic gravitational waves
  using the first science data from the GEO 600 and LIGO detectors}.
\newblock {\em Phys.\ Rev.\ D}, 69:082004, 2004.

\bibitem{Abbott2005a}
B.~Abbott et~al.
\newblock {First all-sky upper limits from LIGO on the strength of periodic
  gravitational waves using the Hough transform}.
\newblock {\em Phys.\ Rev.\ D}, 72(10):102004, 2005.

\bibitem{Abbott2008a}
B.~Abbott et~al.
\newblock {All-sky search for periodic gravitational waves in LIGO S4 data}.
\newblock {\em Phys.\ Rev.\ D}, 77(2):022001, 2008.

\bibitem{Abbott2009c}
B.~Abbott et~al.
\newblock {All-sky LIGO search for periodic gravitational waves in the early
  fifth-science-run data}.
\newblock {\em Phys.\ Rev.\ Lett.}, 102(11):111102, 2009.

\bibitem{Palomba:1999su}
C.~Palomba.
\newblock Pulsars ellipticity revised.
\newblock {\em A\&A}, 354:163--168, 2000.

\bibitem{Keitel:2013wga}
David Keitel, Reinhard Prix, Maria~Alessandra Papa, Paola Leaci, and Maham
  Siddiqi.
\newblock {Search for continuous gravitational waves: improving robustness
  versus instrumental artifacts}.
\newblock {\em Phys.\ Rev.\ D}, 89:064023, 2014.

\bibitem{shaltev:2014}
M.~{Shaltev}, P.~{Leaci}, M.~A. {Papa}, and R.~{Prix}.
\newblock {Fully coherent follow-up of continuous gravitational-wave
  candidates: An application to Einstein@Home results}.
\newblock {\em Phys.\ Rev.\ D}, 89(12):124030, 2014.

\end{thebibliography}

\end{document}